\documentclass[iop,numberedappendix,appendixfloats]{emulateapj}
\usepackage{color}
\usepackage{graphicx}
\usepackage{subfigure}
\usepackage{epsfig}
\usepackage{enumerate}
\usepackage{amsmath}
\usepackage{booktabs}
\usepackage{mathtools}
\usepackage{float}
\usepackage{euscript}

\newcommand{\af}{[$\alpha$/Fe] }
\newcommand{\mh}{[Fe/H] }
\begin{document}
\DeclareGraphicsExtensions{.ps,.pdf,.png,.jpg}
%\DeclareGraphicsRule{.ps}{pdf}{.pdf}{.jpg}
\title{Chemical Cartography with APOGEE: Metallicity Distribution Functions and the Chemical Structure of the Milky Way Disk}
%No G-Dwarfs, No Problem?

%primary
\author{Michael R. Hayden\altaffilmark{1}, 
Jo Bovy\altaffilmark{2,3}, 
Jon A. Holtzman\altaffilmark{1},
David L. Nidever \altaffilmark{4},
Jonathan C. Bird \altaffilmark{5},
David H. Weinberg \altaffilmark{6},
Brett H. Andrews\altaffilmark{7},
%second group
Carlos Allende Prieto\altaffilmark{8,9},
Friedrich Anders\altaffilmark{10},
Timothy C. Beers\altaffilmark{11},
Dmitry Bizyaev\altaffilmark{12},
Cristina Chiappini\altaffilmark{10,13},
Katia Cunha \altaffilmark{14,15},
Peter Frinchaboy \altaffilmark{16},
Domingo~A.~Garc\'{\i}a-Her\'{n}andez\altaffilmark{8,9},
Ana~E.~Garc\'{\i}a~P\'{e}rez\altaffilmark{17,8,9},
L\'{e}o Girardi\altaffilmark{18,13},
Paul Harding\altaffilmark{19},
Fred R. Hearty \altaffilmark{20,17},
Jennifer A. Johnson\altaffilmark{6},
Steven R. Majewski \altaffilmark{17},
Szabolcs M{\'e}sz{\'a}ros\altaffilmark{21},
Ivan Minchev\altaffilmark{10},
Robert O'Connell\altaffilmark{17},
Kaike Pan\altaffilmark{12},
Annie C.Robin\altaffilmark{22},
Ricardo P. Schiavon\altaffilmark{23},
Donald P. Schneider\altaffilmark{20,24},
Mathias Schultheis\altaffilmark{25},
Matthew Shetrone\altaffilmark{26},
Michael Skrutskie\altaffilmark{17},
Matthias Steinmetz\altaffilmark{10},
Verne Smith\altaffilmark{14},
Olga Zamora\altaffilmark{8,9},
Gail Zasowski\altaffilmark{27}
%Carlos Allende Prieto\altaffilmark{4},
%Timothy C. Beers\altaffilmark{5},
%Katia Cunha\altaffilmark{6,7},                    
%Peter M. Frinchaboy\altaffilmark{8},
%Ana E. Garc\'{\i}a P\'erez\altaffilmark{3},
%L\'eo Girardi \altaffilmark{9,10}, 
%Fred R. Hearty \altaffilmark{11},
%Jennifer A. Johnson\altaffilmark{12}, 
%Young Sun Lee\altaffilmark{1},
%David Nidever\altaffilmark{13},
%Ricardo P. Schiavon\altaffilmark{14},
%Katharine J. Schlesinger\altaffilmark{15},
%Donald P. Schneider\altaffilmark{11,16},
%Mathias Schultheis\altaffilmark{17},
%Matthew Shetrone\altaffilmark{18},
%Verne V. Smith\altaffilmark{19,7},
%Gail Zasowski\altaffilmark{20},
%Dmitry Bizyaev\altaffilmark{21},  
%Diane Feuillet\altaffilmark{1},
%Sten Hasselquist\altaffilmark{1},
%Karen  Kinemuchi\altaffilmark{21},
%Elena  Malanushenko\altaffilmark{21}, 
%Viktor Malanushenko\altaffilmark{21},  
%Robert O'Connell\altaffilmark{3},
%Kaike Pan\altaffilmark{21},      
%Keivan Stassun\altaffilmark{22}     
}

\altaffiltext{1}{New Mexico State University, Las Cruces, NM 88003, USA (mrhayden, holtz@nmsu.edu)}

\altaffiltext{2}{Institute for Advanced Study, Einstein Drive, Princeton, NJ 08540, USA (bovy@ias.edu)}

\altaffiltext{3}{Bahcall Fellow}
 
\altaffiltext{4}{Department of Astronomy, University of Michigan, Ann Arbor, MI 48109, USA (dnidever@umich.edu)}

\altaffiltext{5}{Department of Physics and Astronomy, Vanderbilt Universiy, 6301 Stevenson Center, VU Station B \#351807, Nashville, TN 37235, USA (jonathan.bird@vanderbilt.edu)}

\altaffiltext{6}{Department of Astronomy and CCAPP, The Ohio State University, Columbus, OH 43210, USA (dhw, jaj@astronomy.ohio-state.edu)}

\altaffiltext{7}{PITT PACC, Department of Physics and Astronomy, University of Pittsburgh, 3941 O'Hara Street, Pittsburgh, PA 15260, USA (andrewsb@pitt.edu)}

\altaffiltext{8}{%Carlos
  Instituto de Astrof{\'{\i}}sica de Canarias (IAC), E-38200 La Laguna, Tenerife, Spain} %10
\altaffiltext{9}{%Carlos
  Departamento de Astrof{\'{\i}}sica, Universidad de La Laguna (ULL), E-38206 La Laguna, Tenerife, Spain}
\altaffiltext{10}{%Anders, Chiappini, Minchev, Steinmetz
  Leibniz-Institut f\"{u}r Astrophysik Potsdam (AIP), An der Sternwarte 16, 14482, Potsdam, Germany (fanders, cristina.chiappini,msteinmetz@aip.de, iminchev1@gmail.com)
}
\altaffiltext{11}%Beers
{Dept. of Physics and JINA-CEE: Joint Institute for Nuclear Astrophysics--Center for the Evolution of the Elements, University of Notre Dame, Notre Dame, IN 46556, USA  (tbeers@nd.edu)}

\altaffiltext{12}{%Bizyaev,Pan
  Apache Point Observatory and New Mexico State University, P.O. Box 59, Sunspot, NM, 88349-0059, USA (dmbiz, kpan@apo.nmsu.edu)}

\altaffiltext{13}{%Chiappini,Girardi
  Laborat\'{o}rio Interinstitucional de e-Astronomia - LIneA, Rua Gal. Jos\'{e} Cristino 77, Rio de Janeiro, RJ - 20921-400, Brazil}

\altaffiltext{14}{%Cunha,Smith
National Optical Astronomy Observatories, Tucson, AZ 85719, USA (cunha, vsmith@email.noao.edu)}

\altaffiltext{15}{%Cunha
Observat\'orio Nacional, S\~ao Crist\'ov\~ao, Rio de Janeiro, Brazil}

\altaffiltext{16}{%Frinchaboy
  Department of Physics and Astronomy, Texas Christian University, Fort Worth, TX 76129, USA (p.frinchaboy@tcu.edu)}

\altaffiltext{17}{%Garcia-Perez,Majewski, O'Conell, Skrutskie
Dept. of Astronomy, University of Virginia, Charlottesville, VA 22904-4325, USA (aeg4x, srm4n, rwo, mfs4n@virginia.edu)}

\altaffiltext{18}{%Girardi,Rodrigues
  Osservatorio Astronomico di Padova - INAF, Vicolo dell'Osservatorio 5, I-35122 Padova, Italy (leo.girardi@aopd.inaf.it)}

\altaffiltext{19}{%Harding
Department of Astronomy, Case Western Reserve University, 10900 Euclid Ave, Cleveland, OH 44106, USA (paul.harding@case.edu)}

\altaffiltext{20}{%Heart, Schneider
Department of Astronomy and Astrophysics, The Pennsylvania State University, University Park, PA 16802, USA (frh10,dps7@psu.edu)}

\altaffiltext{21}{%Meszaros
ELTE Gothard Astrophysical Observatory, H-9704 Szombathely, Szent Imre herceg st. 112, Hungary (meszi@gothard.hu)}
 
\altaffiltext{22}{%Robin
  Institute Utinam, CNRS UMR6213, Universit\'{e} de Franche-Comt\'{e}, OSU THETA de Franche-Comt\'{e}-Bourgogne, Besancon, France (annie@obs-besancon.fr)}

\altaffiltext{23}{%Schiavon
  Astrophysics Research Institute, IC2, Liverpool Science Park, Liverpool John Moores University, 146 Brownlow Hill, Liverpool, L3 5RF, UK (rpschiavon@gmail.com)}

\altaffiltext{24}{%Schneider
Institute for Gravitation and the Cosmos, The Pennsylvania State University, University Park, PA 16802, USA}

\altaffiltext{25}{%Schultheis
  Laboratoire Lagrange (UMR7293), Universit\'{e} de Nice Sophia Antipolis, CNRS, Observatoire de la C\^{o}te d'Azur, BP 4229, 06304 Nice Cedex 4, France (mathias.schultheis@oca.edu)}

\altaffiltext{26}{%Shetrone
  The University of Texas at Austin, McDonald Observatory, TX 79734, USA (shetrone@astro.as.utexas.edu)}

\altaffiltext{27}{%Zasowski
  Department of Physics and Astronomy, Johns Hopkins University, Baltimore, MD 21218, USA (gail.zasowski@gmail.com)}

\begin{abstract}
Using a sample of 69,919 red giants from the SDSS-III/APOGEE Data Release 12, we measure the distribution of stars in the \af vs. \mh plane and the metallicity distribution functions (MDF) across an unprecedented volume of the Milky Way disk, with radius $3<R<15$ kpc and height $|z|<2$ kpc. Stars in the inner disk ($R<5$ kpc) lie along a single track in \af vs. \mh, starting with $\alpha$-enhanced, metal-poor stars and ending at \af$\sim0$ and [Fe/H]$\sim+0.4$. At larger radii we find two distinct sequences in \af vs. \mh space, with a roughly solar-$\alpha$ sequence that spans a decade in metallicity and a high-$\alpha$ sequence that merges with the low-$\alpha$ sequence at super-solar [Fe/H]. The location of the high-$\alpha$ sequence is nearly constant across the disk, however there are very few high-$\alpha$ stars at $R>11$ kpc. The peak of the midplane MDF shifts to lower metallicity at larger $R$, reflecting the Galactic metallicity gradient. Most strikingly, the shape of the midplane MDF changes systematically with radius, with a negatively skewed distribution at $3<R<7$ kpc, to a roughly Gaussian distribution at the solar annulus, to a positively skewed shape in the outer Galaxy. For stars with $|z|>1$ kpc or \af$>0.18$, the MDF shows little dependence on $R$. The positive skewness of the outer disk MDF may be a signature of radial migration; we show that blurring of stellar populations by orbital eccentricities is not enough to explain the reversal of MDF shape but a simple model of radial migration can do so.

%\vspace{0.3cm}}
\end{abstract}
\section{Introduction} 
The Milky Way is an excellent testing ground of our understanding of Galaxy evolution, due to the ability to resolve individual stars and study stellar populations in greater detail than in other galaxies.  However, understanding the composition, structure, and origin of the Milky Way disk is still currently one of the outstanding questions facing astronomy, and there is great debate about this topic (e.g., \citealt{Rix2013}). The ability to resolve individual stars allows one to trace the fossil record of the Milky Way across the disk, as the stars contain the chemical footprint of the gas from which they formed.

Observations of stars in the Milky Way have led to the discovery of several chemical and kinematic properties of the disk of the Galaxy, such as the thick disk (e.g., \citealt{Yoshii1982,Gilmore1983}), chemical abundance gradients (e.g., \citealt{Hartkopf1982,Cheng2012b,Anders2014,Hayden2014,Schlesinger2014}), and the G-dwarf problem \citep{VandenBergh1962,Pagel1975} from the study of the metallicity distribution function (MDF) of the solar neighborhood \citep{VandenBergh1962,Casagrande2011,Lee2011,Schlesinger2012}.  Much of the previous work on the Milky Way disk has focused on the solar neighborhood, or tracer populations (e.g., Cepheid variables, HII regions) that span a narrow range in age and number only a few hundred objects even in the most thorough studies. The advent of large scale surveys such as SEGUE \citep{Yanny2009}, RAVE \citep{Steinmetz2006}, APOGEE \citep{Majewski2015}, GAIA-ESO \citep{Gilmore2012}, and HERMES-GALAH \citep{Freeman2012} aim to expand observations across the Milky Way and will greatly increase the spatial coverage of the Galaxy with large numbers of stars. In this paper we use observations from the twelfth data release of SDSS-III/APOGEE \citep{Alam2015} to measure the distribution of stars in the \af vs. \mh plane and the metallicity distribution function across the Milky Way galaxy with large numbers of stars over the whole radial range of the disk. [$\alpha$/H] is defined by the elements O, Mg, Si, S, Ca, and Ti changing together with solar proportions.  Using standard chemical abundance bracket notation, \af is [$\alpha$/H]--[Fe/H].

Different stellar populations can be identified in chemical abundance space, with the $\alpha$ abundance of stars separating differing populations. The distribution of stars in the \af vs. \mh plane shows two distinct stellar populations in the solar neighborhood (e.g., \citealt{Fuhrmann1998,Prochaska2000,Reddy2006,Adibekyan2012,Haywood2013,Anders2014,Bensby2014,Nidever2014,Snaith2014a}), with one track having roughly solar-\af ratios across a large range of metallicities, and the other track having a high-\af ratio at low metallicity that is constant with \mh until \mh$\sim-0.5$, at which point there is a knee and the \af ratio decreases at a constant rate as a function of \mh, eventually merging with the solar-\af track at \mh$\sim0.2$ dex. The knee in the high-\af sequence is likely caused by the delay time for the onset of Type Ia SNe (SNeIa): prior to formation of the knee, core collapse supernovae (SNII) are the primary source of metals in the ISM, while after the knee SNeIa begin to contribute metals, enriching the ISM primarily in iron peak elements and lowering the \af ratio.  

Stars on the \af-enhanced track have much larger vertical scale-heights than solar-\af stars (e.g., \citealt{Lee2011,Bovy2012,Bovy2012b,Bovy2012a}) and make up the stellar populations belonging to the thick disk.  \citet{Nidever2014} used the APOGEE Red Clump Catalog \citep{Bovy2014} to analyze the stellar distribution in the \af vs. \mh plane across the Galactic disk, and found that the high-\af (thick disk) sequence was similar over the radial range covered in their analysis ($5<R<11$ kpc). The thick disk stellar populations are in general observed to be more metal-poor, $\alpha$-enhanced, have shorter radial scale-lengths, larger vertical scale-heights, and hotter kinematics than most stars in the solar neighborhood (e.g., \citealt{Bensby2003,AllendePrieto2006,Bensby2011,Bovy2012a,Cheng2012a,Anders2014}), although there do exist thick disk stars with solar-\af abundances and super-solar metallicities \citep{Bensby2003,Bensby2005,Adibekyan2011,Bensby2014,Nidever2014,Snaith2014a}. However, the exact structure of the disk is still unknown, and it is unclear whether the disk is the superposition of multiple components (i.e., a thick and thin disk), or if the disk is a continuous sequence of stellar populations (e.g., \citealt{Ivezic2008,Bovy2012,Bovy2012b,Bovy2012a}), or if the structure varies with location in the Galaxy. 

Meanwhile, \citet{Nidever2014} found that the position of the locus of low-\af (thin disk) stars depends on location within the Galaxy (see also \citealt{Edvardsson1993}), and it is possible that in the inner Galaxy the high- and low-\af populations are connected, rather than distinct. Most previous observations were confined to the solar neighborhood, and use height above the plane or kinematics to separate between thick and thin disk populations. However, kinematical selections often misidentify stars \citep{Bensby2014}, and can remove intermediate or transitional populations; which may bias results \citep{Bovy2012a}.

Observations of the metallicity distribution function (MDF) at different locations in the Galaxy can provide information about the evolutionary history across the disk.  The MDF has generally only been well characterized in the solar neighborhood (e.g., \citealt{VandenBergh1962,Nordstrom2004,Ak2007,Casagrande2011,Siebert2011,Schlesinger2012}) and in the Galactic bulge (\citealt{Zoccali2008,Gonzalez2013,Ness2013}). The first observations of the MDF outside of the solar neighborhood were made using APOGEE observations, and found differences in the MDF as a function of Galactocentric radius \citep{Anders2014}. Metallicity distribution functions have long been used to constrain models of chemical evolution. Early chemical evolution models (e.g., \citealt{Schmidt1963,Pagel1975}) that attempted to explain the observed metal distribution in local G-dwarfs were simple, closed-box systems (no gas inflow or outflow) that over predicted the number of metal-poor stars relative to observations. This result commonly known as the ``G-dwarf problem'' (e.g., \citet{Pagel1975,Rocha-Pinto1996,Schlesinger2012}). Solutions to the G-dwarf problem include gas inflow and outflow (e.g., \citealt{Pagel1997}); observations of the MDF led to the realization that gas dynamics play an important role in the chemical evolution of galaxies. However, it is not clear if the G-dwarf problem exists at all locations in the Galaxy, as there have been limited observations outside of the solar circle. %Observations of the MDF across the disk are critical to our ability to constrain the evolutionary history of the Milky Way, because chemical evolution is not likely to be uniform throughout the disk (e.g., inside-out formation, upside-down formation). 

%Simulations and models of the chemical and kinematic evolution of the Milky Way have become increasingly sophisticated (e.g., \citealt{Hou2000,Chiapinni2001,Schoenrich2009a,Kubrik2013,Minchev2014a}) and attempt to explain both the chemical and dynamical history of the Galaxy. Improved chemical evolution models (e.g., \citealt{Hou2000,Chiapinni2001}). Radial migration (e.g., \citealt{Sellwood2002,Schoenrich2009a}). Inside-out (e.g., \citealt{Larson1976,Kobayashi2011,Bird2013}) and/or upside-down (e.g., \citealt{Bird2013}) disk formation mechanism. More observations across the Milky Way across the disk are critical to constraining these models and improving our understanding of the Milky Way.

Simulations and models of the chemical and kinematical evolution of the Milky Way have become increasingly sophisticated (e.g., \citealt{Hou2000,Chiappini2001,Schonrich2009a,Kubryk2013,Minchev2013}) and attempt to explain both the chemical and dynamical history of the Galaxy. Recent chemical evolution models (e.g., \citealt{Hou2000,Chiappini2001}) treat the chemistry of gas and stars in multiple elements across the entire disk, rather than just the solar neighborhood. Several simulations and models find that ``inside-out'' (e.g., \citealt{Larson1976,Kobayashi2011}) and ``upside-down'' (e.g., \citealt{Bournaud2009,Bird2013}) formation of the Galactic disk reproduces observed trends in the Galaxy such as the radial gradient and the lower vertical scale heights of progressively younger populations. Centrally concentrated hot old disks, as seen in cosmological simulations, would result in a decrease in scale-height with radius, which is not observed. In order to explain the presence of stars at high altitudes in the outer disk, in an inside-out formation scenario, \citet{Minchev2015} suggested that disk flaring of mono-age populations is responsible. Such a view also explains the inversion of metallicity and [alpha/Fe] gradients when the vertical distance from the disk midplane is increased (e.g., \citealt{Boeche2013,Anders2014,Hayden2014}). Alternatively, the larger scale heights of older populations could be a consequence of satellite mergers.% (e.g., \citealt{Minchev2014}).

%and maybe add also: 
%younger populations of stars having lower vertical scale-heights than older populations. 

The radial mixing of gas and stars from their original birth radii has also been proposed as an important process in the evolution of the Milky Way disk (e.g., \citealt{Wielen1996,Sellwood2002,Roskar2008,Schonrich2009a,Loebman2011,Solway2012,Halle2015}). Radial mixing occurs through blurring, in which stars have increasingly more eccentric orbits and therefore variable orbital radii, and churning, where stars experience a change in angular momentum and migrate to new locations while maintaining a circular orbit. However, there is much debate on the relative strength of mixing processes throughout the disk. Recent observations and modeling of the solar neighborhood have suggested that the local chemical structure of the disk can be explained by blurring alone \citep{Snaith2014a}, and that churning is not required, but see \citet{Minchev2014a}. 

To disentangle these multiple processes and characterize the history of the Milky Way disk, it is crucial to map the distribution of elements throughout the disk, beyond the solar neighborhood. This is one of the primary goals of the SDSS-III/APOGEE survey observed 146,000 stars across the Milky Way during three years of operation. APOGEE is a high-resolution (R$\sim$22,500) spectrograph operating in the $H$-band, where extinction is 1/6 that of the $V$ band. This allows observations of stars lying directly in the plane of the Galaxy, giving an unprecedented coverage of the Milky Way disk. The main survey goals were to obtain a uniform sample of giant stars across the disk with high resolution spectroscopy to study the chemical and kinematical structure of the Galaxy, in particular the inner Galaxy where optical surveys cannot observe efficiently due to high extinction. The APOGEE survey provides an RV precision of $\sim100$ m s$^{-1}$ \citep{Nidever2015}, and chemical abundances to within 0.1--0.2 dex for 15 different chemical elements \citep{GarciaPerez2015}, in addition to excellent spatial coverage of the Milky Way from the bulge to the edge of the disk. 

In this paper we present results from the Twelfth Data Release (DR12; \citealt{Alam2015}) of SDSS-III/APOGEE on the distribution of stars in the \af vs. \mh plane and their metallicity distribution functions, across the Milky Way and at a range of heights above the plane. In Section 2 we discuss the APOGEE observations, data processing, and sample selection criteria. In Section 3 we present our observed results for the distribution of stars in \af vs. \mh plane and metallicity distribution functions. In Section 4 we discuss our findings in the context of chemical evolution models. In the Appendix we discuss corrections for biases due to survey targeting, sample selection, population effects, and errors in the \af determination.

\section{Data and Sample Selection}

Data are taken from DR12, which contains 
stellar spectra and derived stellar parameters for stars observed during 
the three years of APOGEE. APOGEE is one of the main SDSS-III surveys \citep{Eisenstein2011}, which uses the SDSS 2.5m telescope \citep{Gunn2006} to obtain spectra for hundreds of stars per exposure. These stars cover a wide spatial extent of the Galaxy, and  span a range of magnitudes between $8<H<13.8$ for primary 
science targets. Target selection is described in detail in the APOGEE targeting paper \citep{Zasowski2013} and the APOGEE DR10 paper \citep{Ahn2014}. Extinction and dereddening for each individual star is determined using the Rayleigh-Jeans Color Excess method (RJCE, \citealt{Majewski2011}),
which uses 2MASS photometry \citep{Skrutskie2006} in conjunction with near-IR photometry from the Spitzer/IRAC \citep{Fazio2004} GLIMPSE surveys \citep{Benjamin2003,Churchwell2009} where available, or from WISE \citep{Wright2010}. In-depth discussion of observing and reduction procedures is described in \citep{Hayden2014}, the DR10 paper \citep{Ahn2014}, the APOGEE reduction pipeline paper \citep{Nidever2015}, the DR12 calibration paper \citep{Holtzman2015}, the APOGEE linelist paper \citep{Zamora2015}, and the APOGEE Stellar Parameters and Chemical Abundances Pipeline (ASPCAP, \citealt{GarciaPerez2015}) paper.

\begin{figure}[t!]
\centering
\includegraphics[width=3.3in]{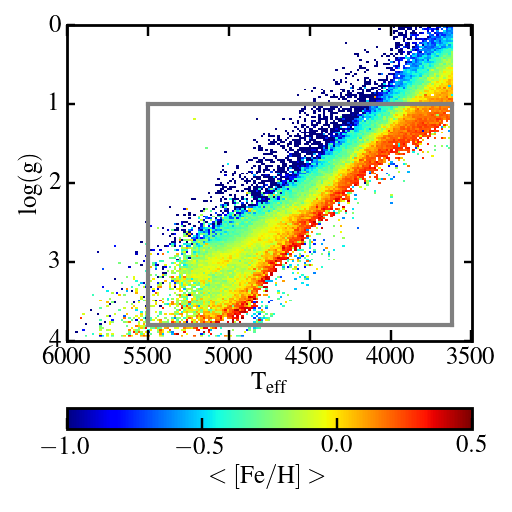}
\caption{The spectroscopic H-R diagram %\citep{Langer2014}
for the full calibrated APOGEE sample, where the mean metallicity in each $\log{g}, T_{\textrm{eff}}$ bin is shown. The gray box denotes the selected sample of 69,919 stars presented in this paper. }%\textbf{Right:} The [$\alpha$/M] vs. \mh plane for the APOGEE DR12 sample. The grey line denotes the separation between low- and high-\af populations used in this paper.}
\label{survey}
\end{figure}

For this paper, we select cool (T$_\textrm{eff}<5500$ K) main survey (e.g., no ancillary program or \textit{Kepler} field) giant stars ($1.0<\log{g}<3.8$) with S/N$>80$. Additionally, stars flagged as ``Bad'' due to being near the spectral library grid edge(s) or having poor spectral fits are removed. The cuts applied to the H-R diagram for DR12 are shown in Figure \ref{survey}. ASPCAP currently has a cutoff temperature of $3500$K on the cool side of the spectral grid (see \citealt{GarciaPerez2015,Zamora2015}), which could potentially bias our results against metal-rich stars. We correct for this metallicity bias by imposing the lower limit on surface gravity of $\log{\textrm{g}}>1.0$, as this mitigates much of the potential bias due to the temperature grid edge in the observed metallicities across the disk; for a detailed discussion see the Appendix. Applying these restrictions to the DR12 catalog, we have a main sample of 69,919 giants within 2 kpc of the midplane of the Milky Way. We restrict our study to the disk of the Milky Way ($R>3$ kpc). For a detailed discussion of the MDF of the Galactic bulge with APOGEE observations see \citet{GarciaPerez2015a}.

\mh has been calibrated to literature values for a large set of reference stars and clusters, and all other spectroscopic parameters are calibrated using clusters to remove trends with temperature, as described in \citet{Holtzman2015}.  These corrections are similar to those applied to the DR10 sample by \citet{Meszaros2013}, as there are slight systematic offsets observed in the ASPCAP parameters compared to reference values, and in some cases trends with other parameters (e.g., abundance trends with effective temperature). The accuracy of the abundances has improved from DR10, in particular for \af, as self-consistent model atmospheres rather than scaled solar atmospheres were used in the latest data release (see \citealt{Zamora2015}), improving the accuracy of many parameters. The typical uncertainties in the spectroscopic parameters from \citet{Holtzman2015} are: 0.11 dex in $\log{g}$, 92 K in T$_{\textrm{eff}}$, and 0.05 dex in \mh and \af. 

Giants are not perfectly representative of underlying stellar populations, as they are evolved stars. There is a bias against the oldest populations when using giants as a tracer population, with the relative population sampling being $\propto\tau^{-0.6}$, where $\tau$ is age (e.g., \citep{Girardi2001}). It is difficult to correct for this population sampling effect, as it depends on the detailed star formation histories at different locations throughout the Galaxy, and likely requires detailed population synthesis models. Because of this, we do not correct for the non-uniform age sampling of giants and our MDFs are slightly biased against the oldest (and potential more metal-poor) stars of the disk. For additional discussion on population sampling, see the Appendix.

%We split the sample into low- and high-\af subsamples, as shown in Figure \ref{alphasplit}. We note that there is a small set of very cool stars (T$_\textrm{eff}<3800$ K that have suspect \af ratios, at \af$\sim+0.2$ and \mh$\sim+0.25$. The higher \af abundance is driven by oxygen lines that effect the coolest stars. 

%\subsection{Low- and High-\af Subsamples}

%Several previous studies have observed differences in derived metallicities, gradients, and kinematics between low- and high-\af populations (e.g. \citealt{Lee2011,Cheng2012a,Bensby2014,Bergemann2014,Hayden2014}). Low-\af populations are often thought of as a thin disk analog, while the high-\af populations are more representative of the thick disk. Motivated by these previous studies, we split the APOGEE DR12 sample using similar formalism to \citet{Lee2011} and \citet{Hayden2014}.

%\begin{figure}[ht!]
%\centering
%\includegraphics[width=3.3in]{alphacuts.png}
%\includegraphics[width=3.5in]{alphacuts.png}
%\caption{The APOGEE DR12 sample in \af vs. \mh space. We split the sample into low- and high-\af subsamples, as denoted by the grey line.}
%\label{alphasplit}
%\end{figure}

%\citet{Santiago2015}
\subsection{Distances}

Distances for each star are determined from the derived stellar parameters and PARSEC isochrones from the Padova-Trieste group \citep{Bressan2012}
based on Bayesian statistics, following methods described by \citet{Burnett2010},\citet{Burnett2011},and \citet{Binney2014}; see also \citet{Santiago2015}. The isochrones range in metallicity from $-2.5<\textrm{\mh}<+0.6$, with a spacing of 0.1 dex, and ages ($\tau$) ranging from 100 Myr to 20 Gyr with spacing of 0.05 dex in $\log{\tau}$. We calculate the probability of all possible distances using the extinction-corrected magnitude (from RJCE, as referenced above), the stellar parameters from ASPCAP, and the PARSEC isochrones using Bayes' theorem:

\begin{equation}\nonumber
  P(model|data) = \frac{P(data|model)P(model)}{P(data)} 
\end{equation}

\noindent where model refers to the isochrone parameters (T$_{\textrm{eff}}$, $\log{g}$, $\tau$, \mh, etc.) and physical location (in our case the distance modulus).  Data refers to the observed spectroscopic and photometric parameters for the star. For our purposes, we are interested in the distance modulus ($\mu$) only, so P(model$|$data) is:

%\begin{equation}\nonumber
%  p(data|model) \propto \prod_j^n \ \ \displaystyle{\exp\left({\frac{-(o_j-I_j)^2}{2\sigma_{o_{\tiny{j}}}^2}}\right)}
%  \label{max}
%\end{equation}

                                                                                                                                                                                                                                                                                \begin{equation}\nonumber
  %P(DM) \propto \int P(I)P(Prior)\exp{\left(\frac{-(DM-m_{o}-M_{I}-A_{\lambda})^2}{2\sigma^2_{m_o}}\right)dI}
  P(model|data) = P(\mu) \propto \int \prod_j^n \ \ \displaystyle{\exp\left({\frac{-(o_j-I_j)^2}{2\sigma_{o_{\tiny{j}}}^2}}\right)}dI
  \label{b1}
\end{equation}

\noindent where o$_j$ is the observed spectroscopic parameter, I$_j$ is the corresponding isochrone parameter, and $\sigma_{o_{j}}$ is the error in the observed spectroscopic parameter. Additional terms can be added if density priors are included, but we did not include density priors for the distances used in this paper; our effective prior is flat in distance modulus. The distance modulus most likely to be correct given the observed parameters and the stellar models is determined by creating a probability distribution function (PDF) of all distance moduli. We use isochrone points within $3\sigma$ of our observed spectroscopic temperature, gravity, and metallicity to compute the distance moduli, where the errors in the observed parameters are given in the data section above. To generate the PDF,  the equation above is integrated over all possible distance moduli, although in practice we use a range of distance moduli between the minimum and maximum magnitudes from the isochrone grid matches to reduce the required computing time.  

The peak, median, or average of the PDF can be used to estimate the most likely distance modulus for a given star. For this paper, we use the median of the PDF to characterize the distance modulus. The error in the distance modulus is given by the variance of the PDF:

\begin{equation}\nonumber
  \sigma_{\mu} = \sqrt{\frac{\int P(\mu)\mu^2d\mu}{\int P(\mu)d\mu}-<\mu>^2}
\end{equation}

\begin{figure}[t!]
\centering
\includegraphics[width=3.3in]{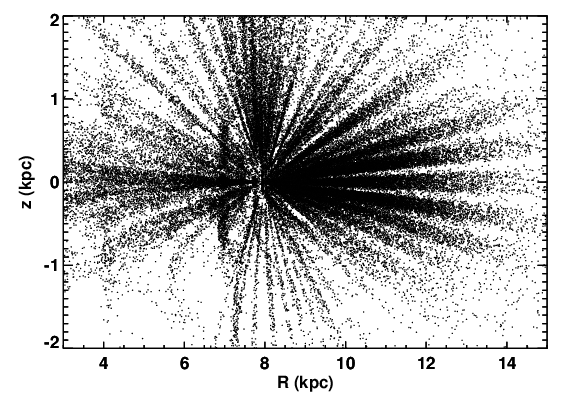}
\caption{The Galactic $R$-$z$ distribution for the sample of 69,919 stars used in this analysis. $R$ is the projected planar distance from the Galactic center, while $z$ is the distance from plane of the Galaxy. Each star is plotted at the location implied by the median of its distance modulus PDF.}
\label{rzmap}
\end{figure}

\noindent The radial distance from the Galactic center is computed assuming a solar distance of 8 kpc from the Galactic center. Distance accuracy was tested by comparing to clusters observed by APOGEE and using simulated observations from TRILEGAL \citep{Girardi2005}. On average, the distances are accurate at the $15-20$\% level. A more detailed discussion of the distances can be found in \citet{Holtzman2015a}. The stellar distribution in the $R$-$z$ plane for the sample used in this paper is shown in Figure \ref{rzmap}.

\section{Results}
\subsection{[$\alpha$/Fe] vs. \mh}

\begin{figure*}[t!]
\centering
\includegraphics[width=6in]{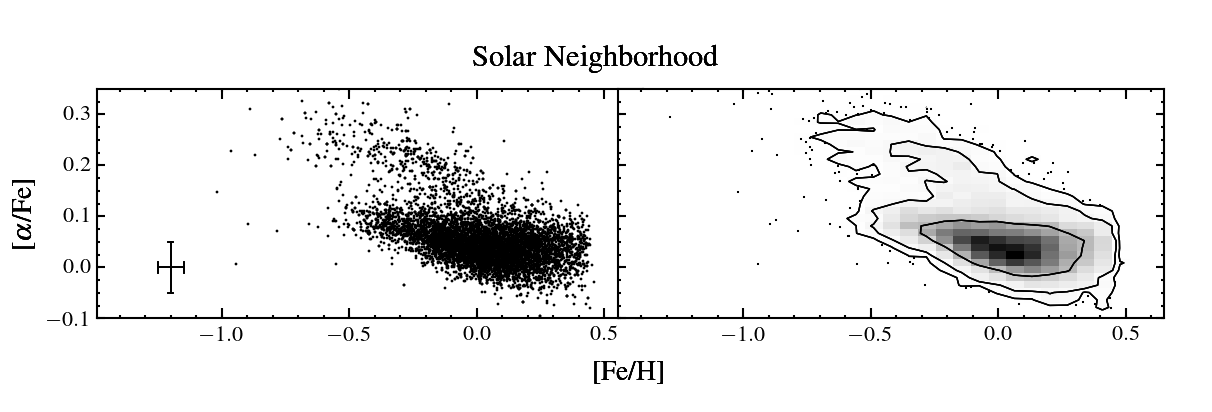}
\caption{The observed [$\alpha$/Fe] vs. \mh distribution for the solar neighborhood ($7<R<9$ kpc, $0<|z|<0.5$ kpc). The left panel is down-sampled and shows only 20\% of the observed data points in the solar circle.  The right panel shows the entire sample in the solar neighborhood, with contours denoting $1,2, 3\sigma$ of the overall densities. There are two sequences in the distribution of stars in the \af vs. \mh plane, one at solar-\af abundances, and one at high-\af abundances that eventually merges with the solar-\af sequence at \mh$\sim0.2$.}
\label{alpham1}
\end{figure*}

\begin{figure*}[ht!]
\centering
\includegraphics[trim=80bp 0 20bp 0,clip,width=7.3in]{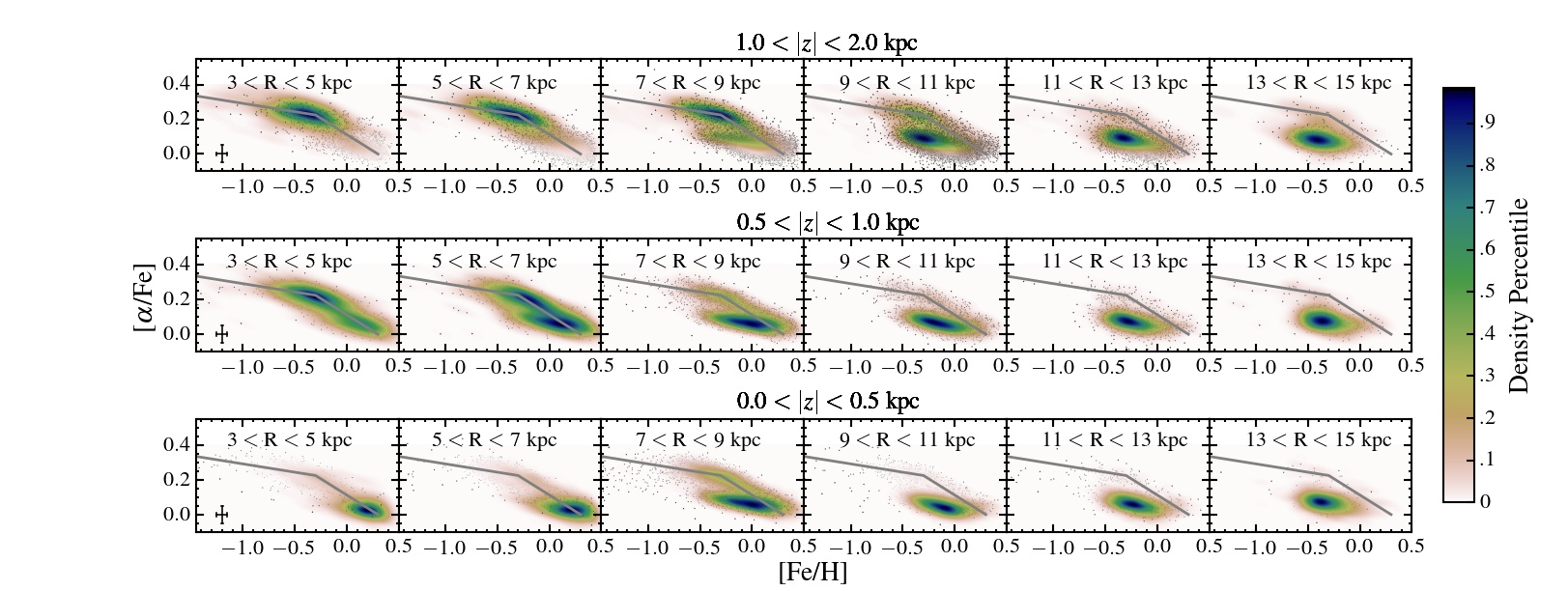}
\caption{The stellar distribution of stars in the \af vs. \mh plane as a function of $R$ and $|z|$. \textbf{Top:} The observed [$\alpha$/Fe] vs. \mh distribution for stars with $1.0<|z|<2.0$ kpc. \textbf{Middle:} The observed [$\alpha$/Fe] vs. \mh distribution for stars with $0.5<|z|<1.0$ kpc. \textbf{Bottom:} The observed [$\alpha$/Fe] vs. \mh distribution for stars with $0.0<|z|<0.5$ kpc. The grey line on each panel is the same, showing the similarity of the shape of the high-\af sequence with $R$. The extended solar-\af sequence observed in the solar neighborhood is not present in the inner disk ($R<5$ kpc), where a single sequence starting at high-\af and low metallicity and ending at solar-\af and high metallicity fits our observations. In the outer disk ($R>11$ kpc), there are very few high-\af stars.}
\label{alphaz}
\end{figure*}

We present results of the distribution of stars in the \af vs. \mh plane in the solar neighborhood ($7<R<9$ kpc, $0<|z|<0.5$ kpc) in Figure \ref{alpham1}. The stellar distribution in the \af vs. \mh plane in the solar neighborhood is characterized by two distinct sequences, one starting at high-\af and the other at approximately solar-\af abundances. We use the term sequences or tracks to describe the behavior of the low and high-\af populations, but this description does not necessarily imply the sequences to be evolutionary in nature. The high-\af sequence has a negative slope, with the \af ratio decreasing as \mh increases, and eventually merges with the low-\af sequence at \mh$\sim+0.2$. The low-\af sequence has a slight decrease in \af abundance as metallicity increases, except for the most metal-rich stars (\mh$>0.2$) where there the trend flattens. The lower envelope of the distribution has a concave-upward, bowl shape. However, these trends are small, and the sequence is within $\sim0.1$ dex in $\alpha$ abundance across nearly a decade in metallicity. It is unclear which sequence the most metal-rich stars belong to in the solar neighborhood, because the low- and high-\af sequences appear to merge at these super-solar metallicities. These observations are similar to previous studies of the solar neighborhood (e.g., \citealt{Adibekyan2012,Ramirez2013,Bensby2014,Nidever2014,Recio-Blanco2014}), which also find two distinct sequences in the \af vs. \mh plane. While some surveys targeted kinematic ``thin'' and ``thick'' disk samples in a way that could amplify bimodality, our sample (and that of \citealt{Adibekyan2011}) has no kinematic selection.

APOGEE allows us to extend observations of the distribution of stars in the \af vs. \mh plane across much of the disk ($3<R<15$ kpc, $|z|<2$ kpc) as shown in Figure \ref{alphaz}). The most striking feature of the stellar distribution in the \af vs. \mh plane in the inner disk ($3<R<5$ kpc) is that the separate low-\af sequence evident in the solar neighborhood is absent--- there appears to be a single sequence starting at low metallicities and high-\af abundances, which ends at approximately solar-\af and high metallicity (\mh$\sim+0.5$). The metal-rich solar-\af stars dominate the overall number density of stars close to the plane. These stars are confined to the midplane, while for $|z|>1$ kpc the majority of the stars in the inner Galaxy are metal-poor with high-\af abundances.

In the $5<R<7$ kpc annulus, the population of low-\af, sub-solar metallicity stars becomes more prominent, revealing the two-sequence structure found in the solar neighborhood. However, the locus of the low alpha sequence is significantly more metal rich ([Fe/H]$\sim0.35$) than our local sample. The mean metallicity of the low alpha sequence and its dependence on radius largely drives the observed Galactic metallicity gradients.  As $|z|$ increases, the relative fraction of high- and low-\af stars changes; for $|z|>1$ kpc, the bulk of the stars belong to the high-\af sequence and have sub-solar metallicities.

%For $5<R<7$ kpc, there are two sequences in the \af vs. \mh plane, similar to the patterns observed in the solar neighborhood, although the locus of the low-\af sequence is significantly more metal-rich reflecting the radial metallicity gradient (e.g., \citealt{Anders2014,Hayden2014}). Close to the plane, the number density of stars is dominated by the low-\af sequence. As $|z|$ increases, the relative fraction of high- and low-\af stars changes; for $|z|>1$ kpc, the bulk of the stars belong to the high-\af sequence and have sub-solar metallicities.

Towards the outer disk ($R>9$ kpc), the locus of the low-\af sequence shifts towards lower metallicities. Much like the rest of the Galaxy, the low-\af sequence dominates the number density close to the plane of the disk. As $|z|$ increases, the high-\af fraction increases (for $9<R<13$ kpc), but it never becomes the dominant population at high $|z|$ as it does in the solar neighborhood or inner disk. For $R>13$ kpc, there are almost no high-\af stars present; at all heights above the plane most stars belong to the low-\af sequence. For $R>11$ kpc, the relative number of super-solar metallicity stars is low compared to the rest of the Galaxy; these stars are confined to the inner regions ($R<11$ kpc) of the disk. The spread in metallicity for the very outer disk ($R>13$ kpc) is small: most stars are within \mh$\sim-0.4\pm0.2$ dex at all heights about the plane. 

\citet{Nidever2014} used the APOGEE Red Clump Catalog \citep{Bovy2014} to categorize the distribution of stars in the \af vs. \mh plane outside of the solar neighborhood. The \citet{Nidever2014} sample is a subset of the same data presented in this paper. The red clump offers more precise distance and abundance determinations compared to the entire DR12 sample, but it covers a more restricted distance and metallicity range. \citet{Nidever2014} find that the high-\af sequence found in the solar neighborhood is similar in shape in all areas of the Galaxy where it could be observed ($5<R<11$ kpc, $0<|z|<2$ kpc). Here we expand these observations to larger distances and find a similar result; the shape of the high-\af sequence does not vary significantly with radius, although the very inner Galaxy does show a hint of small differences. There appears to be a slight shift towards lower-\af for the same metallicities by $~0.05$ dex compared to the high-\af sequence observed in the rest of the disk. This variation may be caused by temperature effects; the stars in the inner Galaxy are all cool (T$_{\textrm{eff}}<4300$ K), and there is a slight temperature dependence of \af abundance for cooler stars, as discussed further in the Appendix. Although the high-\af sequence appears similar at all observed locations, as noted above the number of stars along the high-\af sequence begins to decrease dramatically for $R>11$ kpc: there are almost no stars along the high-\af sequence in the very outer disk ($13<R<15$ kpc). 

To summarize our results for the distribution of stars in the \af vs. \mh plane:

\begin{itemize}
  \item There are two distinct sequences in the solar neighborhood, one at high-\af, and one at solar-\af, which appear to merge at \mh $\sim+0.2$. At sub-solar metallicities there is a distinct gap between these two sequences.
  \item The abundance pattern for the inner Galaxy can be described as a single sequence, starting at low-metallicity and high-\af and ending at approximately solar-\af and \mh $=+0.5$. The most metal-rich stars are confined to the midplane.
  \item The high-\af sequence appears similar at all locations in the Galaxy where it is observed ($3<R<13$ kpc).
  \item Stars with high-\af ratios and the most metal-rich stars (\mh$>0.2$) have spatial densities that are qualitatively consistent with short radial scale-lengths or a truncation at larger radii and have low number density in the outer disk.
  \item The relative fraction of stars between the low- and high-\af sequences varies with disk height and radius.
    
\end{itemize}

%Relative number of high- and low-\af stars at different locations.

\subsection{Metallicity Distribution Functions}

\begin{deluxetable*}{crccccc}
\tabletypesize{\footnotesize}
\tablecolumns{7}
\tablewidth{0pt}
\tablecaption{Metallicity Distribution Functions in the Milky Way \label{tab:tabmdf}}
\tablehead{
\colhead{$R$ Range (kpc)} & \colhead{N*} & \colhead{$<$\mh$>$} & \colhead{Peak \mh} & \colhead{$\sigma_{\textrm{\mh}}$} & \colhead{Skewness} & \colhead{Kurtosis}}
\startdata
\multicolumn{7}{c}{$1.00< |z| < 2.00$}\\
\hline
$ 3 < R < 5 $ & 465 & -0.42 & -0.27 & 0.29 & -0.48$\pm$0.14 & 1.05$\pm$0.28 \\
$ 5 < R < 7 $ & 846 & -0.36 & -0.33 & 0.29 & -0.32$\pm$0.13 & 1.27$\pm$0.26 \\
$ 7 < R < 9 $ & 4136 & -0.31 & -0.27 & 0.28 & -0.53$\pm$0.06 & 1.52$\pm$0.17 \\
$ 9 < R < 11 $ & 1387 & -0.29 & -0.27 & 0.25 & -0.37$\pm$0.13 & 1.69$\pm$0.36 \\
$ 11 < R < 13 $ & 827 & -0.29 & -0.38 & 0.23 & -0.40$\pm$0.21 & 2.58$\pm$0.62 \\
$ 13 < R < 15 $ & 207 & -0.39 & -0.43 & 0.17 & -0.60$\pm$0.73 & 3.84$\pm$3.16 \\
\hline
\multicolumn{7}{c}{$0.50< |z| < 1.00$}\\
\hline
$ 3 < R < 5 $ & 841 & -0.19 & -0.33 & 0.32 & -0.50$\pm$0.11 & 0.50$\pm$0.31 \\
$ 5 < R < 7 $ & 1408 & -0.12 & -0.18 & 0.29 & -0.50$\pm$0.09 & 0.39$\pm$0.35 \\
$ 7 < R < 9 $ & 4997 & -0.10 & -0.02 & 0.25 & -0.49$\pm$0.06 & 0.67$\pm$0.22 \\
$ 9 < R < 11 $ & 3702 & -0.15 & -0.23 & 0.21 & -0.22$\pm$0.10 & 0.99$\pm$0.47 \\
$ 11 < R < 13 $ & 2169 & -0.23 & -0.27 & 0.19 & +0.28$\pm$0.11 & 0.92$\pm$0.39 \\
$ 13 < R < 15 $ & 568 & -0.33 & -0.33 & 0.19 & -0.60$\pm$0.39 & 4.63$\pm$1.50 \\
\hline
\multicolumn{7}{c}{$0.00< |z| < 0.50$}\\
\hline
$ 3 < R < 5 $ & 2410 & +0.08 & +0.23 & 0.24 & -1.68$\pm$0.12 & 4.01$\pm$0.83 \\
$ 5 < R < 7 $ & 5195 & +0.11 & +0.23 & 0.22 & -1.26$\pm$0.08 & 2.53$\pm$0.52 \\
$ 7 < R < 9 $ & 13106 & +0.01 & +0.02 & 0.20 & -0.53$\pm$0.04 & 0.86$\pm$0.26 \\
$ 9 < R < 11 $ & 19930 & -0.11 & -0.12 & 0.19 & -0.02$\pm$0.03 & 0.49$\pm$0.14 \\
$ 11 < R < 13 $ & 6730 & -0.21 & -0.23 & 0.18 & +0.17$\pm$0.06 & 0.79$\pm$0.21 \\
$ 13 < R < 15 $ & 912 & -0.31 & -0.43 & 0.18 & +0.47$\pm$0.13 & 1.00$\pm$0.29 \\
\enddata
\vspace{-0.4cm}
\tablecomments{Statistics for the different MDFs in the across the Milky Way disk.  The kurtosis is defined as the fourth standardized moment-3, such that a normal distribution has a kurtosis of 0.}
\label{tabmdf}
\end{deluxetable*}

\begin{figure}[ht!]
\centering
\includegraphics[width=3.3in]{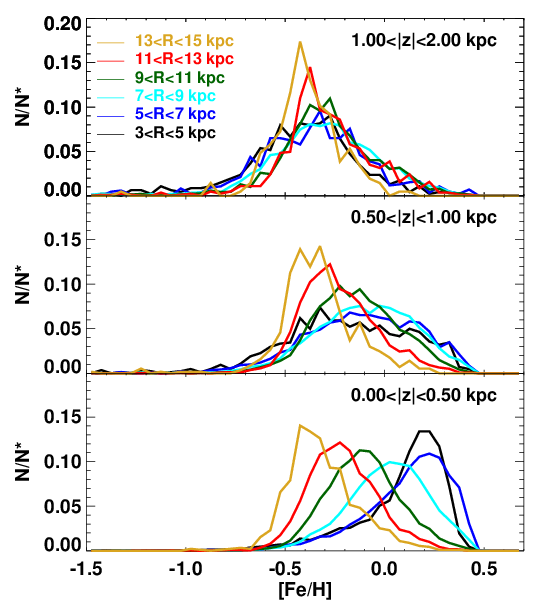}
\caption{The observed MDF for the entire sample as a function of Galactocentric radius over a range of distances from the plane. The shape and skewness is a function of radius and height. Close to the plane, the inner Galaxy ($3<R<5$ kpc) is a negatively skewed distribution and a peak metallicity at $\sim0.25$ dex, while the outer disk ($R>11$ kpc) has a positively skewed distribution with peak metallicity of $\sim-0.4$ dex . For $|z|>1$ kpc, the MDF is fairly uniform at across all radii.\\}
\label{mdf1}
\end{figure}

\begin{figure}[ht!]
\centering
\includegraphics[width=3.3in]{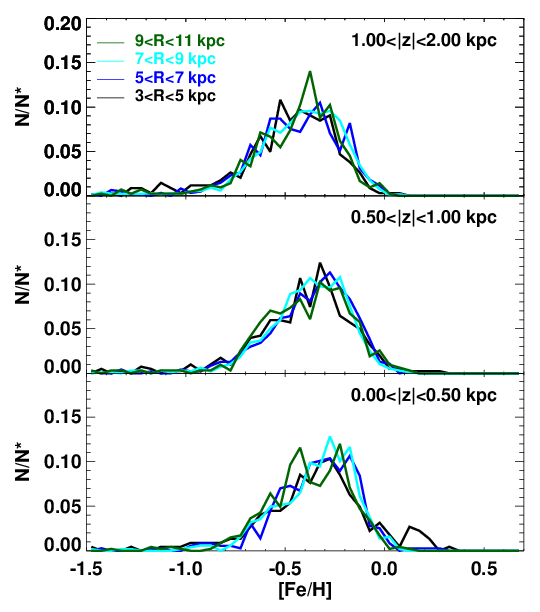}
\caption{The observed MDF for stars with \af$>0.18$ as a function of Galactocentric radius over a range of distances from the plane. There is little variation in the MDF with Galactocentric radius. There is a shallow negative vertical gradient, as stars with $|z|>1$ kpc are slightly more metal-poor than stars close to the plane.\\}
\label{highmdf}
\end{figure}

\begin{deluxetable*}{crccccc}
\tabletypesize{\footnotesize}
\tablecolumns{7}
\tablewidth{0pt}
\tablecaption{$\alpha$ Distribution Functions in the Milky Way \label{tab:tabadf}}
\tablehead{
\colhead{$R$ Range (kpc)} & \colhead{N*} & \colhead{$<[\alpha$/H]$>$} & \colhead{Peak [$\alpha$/H]} & \colhead{$\sigma_{[\alpha/H]}$} & \colhead{Skewness} & \colhead{Kurtosis}}
\startdata
\multicolumn{7}{c}{$1.00< |z| < 2.00$}\\
\hline
$ 3 < R < 5 $ & 468 & -0.20 & -0.08 & 0.26 & -1.12$\pm$0.23 & 3.02$\pm$0.91 \\
$ 5 < R < 7 $ & 853 & -0.16 & -0.08 & 0.27 & -1.25$\pm$0.18 & 4.17$\pm$0.67 \\
$ 7 < R < 9 $ & 4145 & -0.14 & -0.08 & 0.24 & -1.11$\pm$0.09 & 3.70$\pm$0.34 \\
$ 9 < R < 11 $ & 1393 & -0.16 & -0.18 & 0.23 & -0.93$\pm$0.22 & 4.68$\pm$0.90 \\
$ 11 < R < 13 $ & 831 & -0.20 & -0.27 & 0.22 & -0.82$\pm$0.29 & 4.61$\pm$1.06 \\
$ 13 < R < 15 $ & 210 & -0.30 & -0.38 & 0.21 & -1.91$\pm$0.57 & 9.19$\pm$1.80 \\
\hline
\multicolumn{7}{c}{$0.50< |z| < 1.00$}\\
\hline
$ 3 < R < 5 $ & 844 & -0.03 & +0.02 & 0.27 & -1.12$\pm$0.20 & 3.34$\pm$0.87 \\
$ 5 < R < 7 $ & 1409 & +0.02 & +0.07 & 0.23 & -0.73$\pm$0.15 & 1.73$\pm$0.74 \\
$ 7 < R < 9 $ & 4997 & -0.01 & +0.02 & 0.21 & -0.39$\pm$0.06 & 0.78$\pm$0.25 \\
$ 9 < R < 11 $ & 3703 & -0.08 & -0.12 & 0.19 & -0.05$\pm$0.10 & 0.98$\pm$0.48 \\
$ 11 < R < 13 $ & 2170 & -0.16 & -0.23 & 0.18 & 0.34$\pm$0.19 & 1.74$\pm$1.00 \\
$ 13 < R < 15 $ & 568 & -0.25 & -0.27 & 0.17 & -0.42$\pm$0.45 & 4.65$\pm$1.88 \\
\hline
\multicolumn{7}{c}{$0.00< |z| < 0.50$}\\
\hline
$ 3 < R < 5 $ & 2414 & +0.16 & +0.27 & 0.20 & -1.91$\pm$0.20 & 7.17$\pm$1.53 \\
$ 5 < R < 7 $ & 5195 & +0.16 & +0.23 & 0.19 & -1.07$\pm$0.10 & 2.54$\pm$0.78 \\
$ 7 < R < 9 $ & 13109 & +0.06 & +0.07 & 0.18 & -0.34$\pm$0.07 & 1.20$\pm$0.50 \\
$ 9 < R < 11 $ & 19930 & -0.07 & -0.08 & 0.17 & +0.27$\pm$0.03 & 0.35$\pm$0.11 \\
$ 11 < R < 13 $ & 6731 & -0.16 & -0.23 & 0.16 & +0.38$\pm$0.07 & 1.17$\pm$0.37 \\
$ 13 < R < 15 $ & 914 & -0.25 & -0.33 & 0.17 & +0.04$\pm$0.39 & 3.75$\pm$1.82 \\
\enddata
\vspace{-0.4cm}
\tablecomments{Statistics for the different ADFs in the across the Milky Way disk.  The kurtosis is defined as the fourth standardized moment-3, such that a normal distribution has a kurtosis of 0.}
\label{tabadf}
\end{deluxetable*}

\begin{figure}[ht!]
\centering
\includegraphics[width=3.3in]{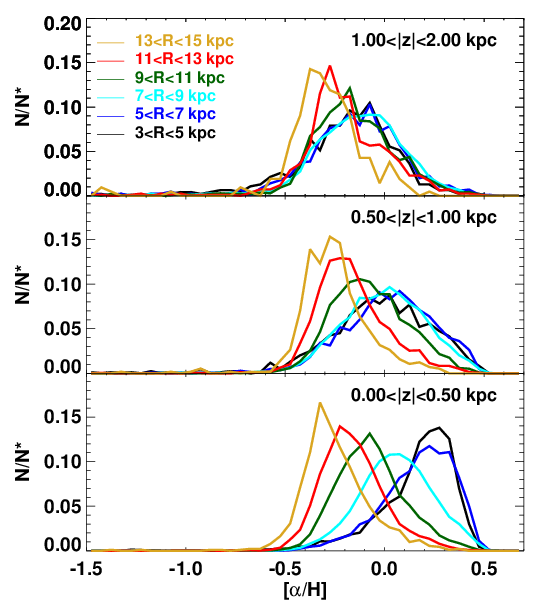}
\caption{The observed distribution of [$\alpha$/H] for the entire sample as a function of Galactocentric radius over a range of distances from the plane. The results are quite similar to that of the MDF, except for stars with $|z|>1$ kpc and $R<9$ kpc, where the ADF has larger abundances than the MDF at the same locations.\\}
\label{amdf}
\end{figure}

With three years of observations, there are  sufficient numbers of stars in each Galactic zone to measure MDFs in a number of radial bins and at different heights above the plane.  We present the MDFs in radial bins of 2 kpc between $3<R<15$ kpc, and at a range of heights above the plane between $0<|z|<2$ kpc, in Figure \ref{mdf1}. The MDFs are computed with bins of 0.05 dex in \mh for each zone. Splitting the sample into vertical and radial bins allows us to analyze the changes in the MDF across the Galaxy, but also minimizes selection effects due to the volume sampling of the APOGEE lines of sight and our target selection. 

Close to the plane (top panel of Figure \ref{mdf1}, $|z|<0.5$ kpc) radial gradients are evident throughout the disk. The peak of the MDF is centered at high metallicities in the inner Galaxy (\mh$=0.32$ for $3<R<5$ kpc), roughly solar in the solar neighborhood (M/H]=$+0.02$ for $7<R<9$ kpc),  and low metallicities in the outer disk (\mh$=-0.48$ for $13<R<15$ kpc). The radial gradients observed in the MDF are similar to those measured across the disk in previous studies with APOGEE, as is the shift in the peak of the MDF (e.g., \citealt{Anders2014,Hayden2014}).% The MDF for $3<R<5$ kpc is remarkably similar to the MDF for $5<R<7$ kpc.

The most striking feature of the MDF close to the plane is the change in shape with radius. The inner disk has a large negative skewness ($-1.68\pm0.12$ for $3<R<5$ kpc, see Table \ref{tabmdf}), with a tail towards low metallicities, while the solar neighborhood is more Gaussian in shape with a slight negative skewness ($-0.53\pm0.04$), and the outer disk is positively skewed with a tail towards high metallicities ($+0.47\pm0.13$ for $13<R<15$ kpc). The shape of the observed MDF of the solar neighborhood is in good agreement with the MDF measured by the GCS \citep{Nordstrom2004,Holmberg2007,Casagrande2011}, who measure a peak of just below solar metallicity and also find a similar negative skewness to our observations. There is a slight offset in the peak metallicity, with the APOGEE observations being more metal-rich by $\sim0.1$ dex, but the shapes are extremely similar.  Close to the plane, the distributions are all leptokurtic, with the inner Galaxy ($3<R<7$ kpc) being more strongly peaked than the rest of the disk.

As $|z|$ increases, the MDF exhibits less variation with radius. For $|z|>1$ kpc (bottom panel of Figure \ref{mdf1}), the MDF is uniform with a roughly Gaussian shape across all radii, although it is more strongly peaked for the very outer disk ($R>13$ kpc). However, the populations comprising the MDF(s) at these heights are not the same. In the inner disk, at large heights above the plane the high-\af sequence dominates the number density of stars. In the outer disk ($R>11$ kpc), the stars above the plane are predominantly solar-\af abundance. The uniformity of the MDF at these large heights is surprising given the systematic change in the \af of the populations contributing to the MDF. The MDF is similar for all heights above the plane in the outer disk ($R>11$ kpc). At these larger heights above the plane, the MDFs are leptokurtic as well, but the trend with radius is reversed compared to the distributions close to the plane. For $|z|>1$ kpc, the distributions in the outer disk ($R>11$ kpc) are more strongly peaked than the MDFs for the rest of the disk.

As noted above, the high-\af sequence is fairly constant in shape with radius. The MDF for stars with high-\af abundance (\af$>0.18$) is presented in Figure \ref{highmdf}. The high-\af sequence appears uniform across the radial range in which it is observed ($3<R<11$ kpc); but does display variation with height above the plane. Close to the plane, the MDF is peaked at \mh$=-0.3$ over the entire radial extent where there are large numbers of high-\af stars; and the shape is also the same at all locations. However, as $|z|$ increases, the MDF shifts to slightly lower metallicities, with a peak \mh$=-0.45$ for $|z|>1$ kpc.  At any given height, there is little variation in the shape or peak of the MDF with radius for these high-\af stars.

%The $\alpha$ elements are primarily produced in SNII, and are a better match to the instantaneous recycling prescriptions used in many simple chemical evolution models. 
Simple chemical evolution models often use instantaneous recycling approximations where metals are immediately returned to the gas reservoir after star formation occurs. This approach may not be a good approximation for all chemical elements in the stellar population, but it is more accurate for $\alpha$ elements, which are produced primarily in SNII. We present the distribution of [$\alpha$/H] (ADF) across the disk in Figure \ref{amdf}. The ADF is similar in appearance to the MDF, with radial gradients across the disk and the most $\alpha$-rich stars belonging to the inner Galaxy. The observed change in skewness with radius of the ADF  (Table \ref{tabadf}) is similar to that of the MDF (Table \ref{tabmdf}). Close to the plane, the ADF of the inner disk ($3<R<5$ kpc) is more strongly peaked than the MDF in the same zone, with a significantly larger kurtosis. The main difference between the ADF and the MDF is above the plane of the disk, we observe radial gradients in the ADF at all heights above the plane, which is not the case for the MDF. At large $|z|$, the ADF is significantly more positive than the MDF from the inner Galaxy out to the solar neighborhood by $\sim0.25$ dex. This result does not hold true in the outer disk above the plane ($R>11$ kpc, $|z|>1$ kpc), where the ADF again has similar abundance trends to that of the MDF.

To summarize our results for the MDFs across the disk:

\begin{itemize}
  \item Metallicity gradients are clearly evident in the MDFs, with the most metal-rich populations in the inner Galaxy.
  \item The shape and skewness of the MDF in the midplane is strongly dependent on location in the Galaxy: the inner disk has a large negative skewness, the solar neighborhood MDF is roughly Gaussian, and the outer disk has a positive skewness.
  %\item The shape of the MDF of the inner disk is similar to those of simple, closed-box chemical evolution models
  \item The MDF becomes more uniform with height. For stars with $|z|>1$ kpc it is roughly Gaussian with a peak metallicity of \mh$\sim-0.4$ across the entire radial range covered by this study ($3<R<15$ kpc).
  \item The MDF for the outer disk ($R>11$ kpc)is uniform at all heights above the plane, showing less variation in metallicity with $|z|$ than the rest of the disk.
  \item The MDF for stars with \af$>0.18$ is uniform with $R$, but has a slight negative vertical gradient.
  \item The ADF has many of the same features observed in the MDF, but shows differences for stars out of the plane ($|z|>1$ kpc) and with $R<9$ kpc, where stars tend to have higher [$\alpha$/H] than \mh.
\end{itemize}

\section{Discussion}

Consistent with previous studies, we find that:
(a) the solar-neighborhood MDF is approximately Gaussian in [Fe/H],
with a peak near solar metallicity,
(b) the distribution of stars in the \af-\mh plane is bimodal,
with a high-\af sequence and a low-\af sequence,
(c) the fraction of stars with high \af increases with $|z|$, and
(d) there is a radial gradient in the mean or median value of \mh
for stars near the midplane.
Like \cite{Nidever2014}, we find that the location of the
high-\af sequence is nearly independent of radius and height
above the plane, a result we are able to extend to larger and smaller $R$
and to lower metallicity.  Two striking new results of this study are
(e) that the \af-\mh distribution of the inner disk
($3<R<5$ kpc) is consistent with a single evolutionary track,
terminating at \mh $\sim+0.4$ and roughly solar \af, and
(f) that the midplane MDF changes shape, from strong negative
skewness at $3<R<7$ kpc to strong positive skewness at
$11<R<15$ kpc, with the solar annulus lying at the transition
between these regimes.

The midplane inner disk MDF has the characteristic shape predicted by
one-zone chemical evolution models, with most stars formed after the
ISM has been enriched to an ``equilibrium'' metallicity controlled
mainly by the outflow mass loading parameter $\eta$
(see \citealt{Andrews2015}, for detailed discussion).
Traditional closed box or leaky box models
(see, e.g., \S 5.3 of \citealt{Binney1998}) are a limiting
case of such models, with no accretion.  These models predict
a metallicity distribution $dN/d\ln Z \propto Z \exp(-Z/p_{\rm eff})$
where the effective yield is related to the IMF-averaged population
yield by $p_{\rm eff}=p/(1+\eta)$.  The positively skewed MDFs of
the outer disk could be a distinctive signature of radial migration,
with a high-metallicity tail populated by stars that were born
in the inner Galaxy.  The change of \af distributions with height
could be a consequence of heating of the older stellar populations
or of forming stars in progressively thinner, ``cooler'' populations
as turbulence of the early star-forming disk decreases.
As discussed by \cite{Nidever2014}, the constancy of the high-\af
sequence implies uniformity of the star formation efficiency and
outflow mass loading during the formation of this population.

In the next subsection we discuss the qualitative comparison
between our results and several recent models of Milky Way chemical
evolution.  We then turn to a more detailed discussion of radial
migration with the aid of simple quantitative models.  We conclude
with a brief discussion of vertical evolution.

\subsection{Comparison to Chemical Evolution Models}

Metallicity distribution functions are useful observational tools in constraining the chemical history of the Milky Way. The first chemical evolution models were simple closed-box systems, with no gas inflow or outflow, and often employed approximations such as instantaneous recycling. These models over-predicted the number of metal-poor stars in the solar-neighborhood compared to observations of G-dwarfs (e.g., \citealt{Schmidt1963,Pagel1975}), a discrepancy known as the ``G-Dwarf Problem''. These first observations made it clear that the chemical evolution of the solar neighborhood could not be described by a simple closed-box model; inflow and outflow of gas, along with more realistic yields (i.e., no instantaneous recycling) were required to reproduce observations. The MDF can therefore be used to inform and tune chemical evolution models and provide information such as the star formation history and relative gas accretion or outflow rates at every location where the MDF can be measured. APOGEE observations provide the first thorough characterizations of the MDF of the disk outside of the solar neighborhood, allowing a more complete characterization of the chemical evolutionary history of the Galaxy.

%%%While the closed-box models did a poor job reproducing the MDF of the solar neighborhood, we note that the overall shape and skew of the MDF in the inner Galaxy is similar to that predicted by these models (e.g., \citealt{Pagel1997,Kirby2011}). The chemical enrichment history of the inner Galaxy is very different than that of the solar neighborhood or the outer disk. \citet{Snaith2014a} split the HARPS sample \citep{Adibekyan2012} of nearby FGK stars into inner and outer disk populations, and fit chemical evolution models to the distributions in \af vs. \mh space. They find that the inner disk is well described by a simple closed-box chemical evolution model, and large inflow and outflow of gas are not required to reproduce their observations. Qualitatively, their results for the inner disk are in excellent agreement with our in-situ inner disk population.

Additions such as gas inflow and outflow to chemical evolution models have been able to better reproduce observations of the solar neighborhood, in particular the MDF and stellar distribution \af vs. \mh plane. The two-infall model from \citet{Chiappini1997,Chiappini2001} treats the disk as a series of annuli, into which gas accretes.  In this model, an initial gas reservoir forms the thick disk, following the high-\af evolutionary track. As the gas reservoir becomes depleted, a lull in star formation occurs. SNeIa gradually lower the \af ratio of the remaining gas reservoir as a second, more gradual infall of pristine gas dilutes the reservoir, lowering the overall metallicity but retaining the low-\af abundance of the ISM. Once the surface density of the gas is high enough star formation resumes, forming the metal-poor end of the solar-\af sequence. The MDF from the two-infall model is in general agreement with our observations of the solar neighborhood (see Figure 7 of \citealt{Chiappini1997}), with a peak metallicity near solar and a slight negative skewness towards lower metallicities. Additionally, this model reproduces general trends found in the distribution of stars in the \af vs. \mh plane, in particular with the dilution of the metallicity of the existing gas reservoir with pristine gas to form the low-\af sequence.  This is one possible explanation for the observed low-\af sequence: pristine gas accretes onto the disk and mixes with enriched gas from the inner Galaxy, keeping solar-\af (or intermediate \af that are later lowered to solar-\af ratios by SNeIa) ratios but lowering the metallicity from +0.5 dex in the inner disk to the lower metallicities found in the outer Galaxy. This model, and its ability to explain phenomenologically both the MDF and stellar distribution in the \af vs. \mh plane of the solar neighborhood, highlight the potential importance of gas flow in the evolutionary history of the Galaxy. %The model of \citet{Snaith2014a} shares some features of the two infall model, but it emphasizes distinctions between the inner and outer disks.

The chemical evolution model from \citet{Schonrich2009a} includes radial migration in the processes governing the evolution of the disk. Their models find that the peak of the MDF is a strong function of radius, with the inner Galaxy being more metal-rich than the outer Galaxy. The peaks of their MDFs are similar to those observed in APOGEE at different radii (see Figure 11 of \citealt{Schonrich2009a}). However, their distributions are significantly more Gaussian and less skewed than we observe in our sample. The model distributions from \citet{Schonrich2009a} appear to have a shift from negative to positive skewness from the inner Galaxy to the outer Galaxy, as we observe, but the magnitude of the skewness is not large. The \citet{Schonrich2009a} results for [O/Fe] vs. [Fe/H] is not in good agreement to the distribution of stars in the \af vs. \mh plane observed with the APOGEE sample (see Figure 9 of \citealt{Schonrich2009a}). Their models do not have two distinct sequences in this plane, but a more continuous distribution of populations that start at low-metallicity with high-\af ratios and end at high-metallicity and solar-\af, similar to many other models with inside-out formation. Their model has have a much larger dynamic range than the APOGEE sample in \af abundance, with their [O/Fe] ratio extending to $\sim+0.6$, while the APOGEE \af abundances extend only to $\sim+0.3$, with the thin disk sequence in APOGEE shifted to slightly higher metallicities and lower-\af abundances than the thin-disk sequence presented by \citet{Schonrich2009a}.
%\citet{Kubryk2013}

Recent N-body smooth particle hydrodynamics simulations from \citet{Kubryk2013} track the impact of migration on the stellar distribution of their disc galaxy. Their MDFs (see Figure 10 of \citealt{Kubryk2013}) are fairly uniform throughout the disk, appearing similar to the MDF of the solar neighborhood in the APOGEE observations, with a peak near solar abundances and a negative skewness. \citet{Kubryk2013} do not find significant shifts in the peak or skewness with radius in their simulations, contrary to what is observed in the APOGEE observations. They postulate that the uniformity of the peak of the MDF is due to the lack of gas infall in the simulation, highlighting the importance that gas dynamics can play in the MDF across the disk.

The most sophisticated chemical evolution models to date use cosmological simulations of a Milky Way analog Galaxy, and paint a chemical evolution model on top of the simulated galaxy. Recent simulations by \citet{Minchev2013} match many observed properties of the disk, such as the stellar distribution in the \af vs. \mh plane (see Figure 12 of \citealt{Minchev2013}),  and the flattening of the radial gradient with height (e.g., \citealt{Anders2014,Hayden2014}, see Figure 10 of \citealt{Minchev2014a}).  Their simulations also provide detailed metallicity distributions for a large range of radii. We find that the MDFs from the simulations closely match the APOGEE observations of the MDF in the solar neighborhood and the uniformity of the MDF above the plane (see Figure 9 of \citealt{Minchev2014a}).  While we have good agreement with the MDFs from \citet{Minchev2014a} above the plane at all radii, the MDFs from their simulation do not reproduce the change in the peak and skewness of the MDF observed close to the plane in the inner and outer disk. The MDFs presented in \citet{Minchev2014a} have the same peak metallicity at all radii, which is not observed in APOGEE close to the plane. The metal-rich components of the MDFs from the simulation are in the wings of the distributions, leading to positively skewed MDFs in the inner Galaxy, and roughly Gaussian shapes in the outer disk. The APOGEE observations have the opposite behavior, with negatively skewed distributions in the inner Galaxy and positively skewed distributions in the outer disk. For this paper, we did not correct for the APOGEE selection function. In the future, we plan to do a more detailed comparison between APOGEE observations and the simulation from \citet{Minchev2013}, in which the selection function is taken into account.

%Maybe compare to Kobayashi2011?

\subsection{Radial Mixing}
Simple chemical evolution models (closed or leaky box) are unable to produce the positively skewed MDFs that we observe in the outer disk. Models that include radial mixing (e.g.,\citealt{Schonrich2009a}) are able to at least produce a more Gaussian-shaped MDF across the disk. The fraction of stars that undergo radial migration is difficult to predict from first principles because it depends in detail on spiral structure, bar perturbations, and perturbations by and mergers with satellites (e.g., \cite{Roskar2008,Bird2012,Bird2013,Minchev2013}).% Some mixing of stars born at different Galactocentric radii occurs inevitably from the ``blurring'' of populations with eccentric orbits. The term ``radial migration'' is usually associated more specifically with ``churning,' a term introduced by \cite{Schonrich2009a} to describe the change of guiding center radii induced by the interaction of stellar orbits with transient spiral arms at co-rotation resonance \citep{Sellwood2002} or by interactions with bar perturbations \citep{Minchev2010,Minchev2011}.

%However, the number of stars that experience radial migration is currently a matter of debate, with some recent results suggesting that observations of the solar neighborhood can be explained entirely by blurring effects (e.g., \citealt{Haywood2013,Snaith2014a}). 

To test the effects of blurring and churning on our observed MDFs, we create a simple model of the MDF across the disk. We assume that the intrinsic shape of the MDF is uniform across the disk, and that the observed change in skewness with radius is due to mixing of populations from different initial birth radii. The disk is modeled as a Dehnen distribution function \citep{Dehnen1999} with a velocity dispersion of $31.4$ km s$^{-1}$, a radial scale length of 3 kpc, and a flat radial velocity dispersion profile. These distribution function parameters adequately fit the kinematics of the main APOGEE sample \citep{BovyVcirc}. We model the initial MDF as a skew-normal distribution with a peak at $+0.4$ dex in the inner Galaxy, a dispersion of $0.1$ dex, and a skewness of $-4$; we assume a radial gradient of $-0.1$ dex kpc$^{-1}$ to shift the peak of the MDFs as a function of radius, keeping the dispersion and skewness fixed.  We then determine the distribution of guiding radii with blurring or churning to determine the effect on the observed MDFs.

%\bibitem[Bovy et al.(2012)]{BovyVcirc} Bovy, J., Allende Prieto, C., Beers, T.~C., et al.\ 2012, \apj, 759, 131 

\subsubsection{Blurring}

\begin{figure}[ht!]
\centering
\includegraphics[width=3.0in]{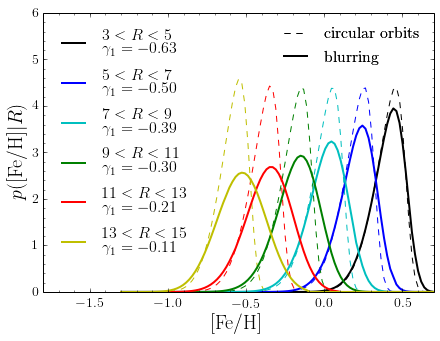}
\caption{The MDF as a function of $R$ for our simple blurring model. The dashed lines show the initial MDFs (which the stars on circular orbits would have) and the solid lines represent the MDFs including the effect of blurring. The skewness of the blurred MDFs are indicated. The magnitude of the skewness diminishes with radius, but it does not change sign.}
\label{blur}
\end{figure}

To determine the effect of blurring on the observed MDF, we use the
model as described above and compute the distribution of guiding radii
$R_{g}$ due to blurring. Assuming a flat rotation curve and
axisymmetry, $R_g\, V_c$ is equal to $R\,V_{\mathrm{rot}}$, where
$V_{\mathrm{rot}}$ is the rotational velocity in the Galactocentric
frame. The blurring distribution $p_b(R_g|R)$ is given by
\begin{equation}\nonumber
  p_b(R_g|R) \sim p(V_{\mathrm{rot}} = \frac{R_g}{R}V_c|R)\,.
  %\label{guiddist}
\end{equation}
The probability $p(V_{\mathrm{rot}}|R)$ can be evaluated using the
assumed Dehnen distribution function. The resulting MDF is 
\begin{equation}\nonumber
  p([\mathrm{Fe/H}]|R) = \int \mathrm{d}R_g\,p([\mathrm{Fe/H}]|R_g)\,p_b(R_g|R)\,
\end{equation}
In Figure \ref{blur} we compare the initial MDF and the MDF with the effects of blurring included. While blurring does reduce
the observed skewness of the MDFs, the MDFs are still negatively skewed at
all radii. This model is simplistic and it is possible that our
underlying assumption regarding the intrinsic shape of the MDF may not
be correct. However, because the intrinsic MDF is unlikely to have positive
skewness anywhere in the Galaxy, it appears that blurring alone is unable
to reproduce the change in sign of the MDF skewness seen in the APOGEE
observations. \citet{Snaith2014a} find that solar neighborhood observations can be adequately explained by the blurring of inner and outer disk populations, but we conclude that such a model cannot explain the full trends with Galactocentric radius measured by APOGEE.

\subsubsection{Radial Migration}

We expand our simple model to include churning to determine if radial
migration is better able to reproduce the observations. We model the
effect of churning on the guiding radii by assuming a diffusion of
initial guiding radii $R_{g,i}$ to final guiding radii $R_{g,f}$, for
stars of age ($\tau$), given by
\begin{equation}\nonumber %might need \usepackage{amsmath}
\begin{split}
  p(& R_{g,f}|R_{g,i},\tau) = \\
  & \mathcal{N}\left(R_{g,f}|R_{g,i},0.01+0.2\,\tau\,R_{g,i}\,e^{-(R_{g,i}-8\,\mathrm{kpc})^2/16\,\mathrm{kpc}^2}\right)\,,
\end{split}
\end{equation}
where $\mathcal{N}(\cdot|m,V)$ is a Gaussian with mean $m$ and
variance $V$. The spread increases as the square root of time and the
largest spread in guiding radius occurs around 8 kpc.  Analysis of
numerical simulations has found that migration can be accurately
described as a Gaussian diffusion
(e.g., \citealt{Brunetti2011,Kubryk2014,VeraCiro2014}); we use the simple
analytic form above to approximate the effect of churning seen in
these more realistic simulations.

%\bibitem[Vera-Ciro et al.(2014)]{VeraCiro14a} Vera-Ciro, C., D'Onghia, E., Navarro, J., \& Abadi, M.\ 2014, \apj, 794, 173 

To obtain the churning distribution $p_c(R_g|R,\tau)$ of initial $R_g$
at a given radius $R$ and age $\tau$, we need to convolve
$p(R_{g,f}|R_{g,i},\tau)$ with the blurring distribution $p_b(R_g|R)$
above
\begin{equation}
  p_c(R_g|R,\tau) = \int \mathrm{d}R_{g,f} p_b(R_{g,f}|R)\,p(R_{g,f}|R_g,\tau)\,,
\end{equation}
where we have used the fact that churning leaves the total surface
density approximately unchanged and therefore that
$p(R_{g}|R_{g,f},\tau) \approx p(R_{g,f}|R_{g},\tau)$.

\begin{figure*}[ht!]
\centering
\includegraphics[width=6.5in]{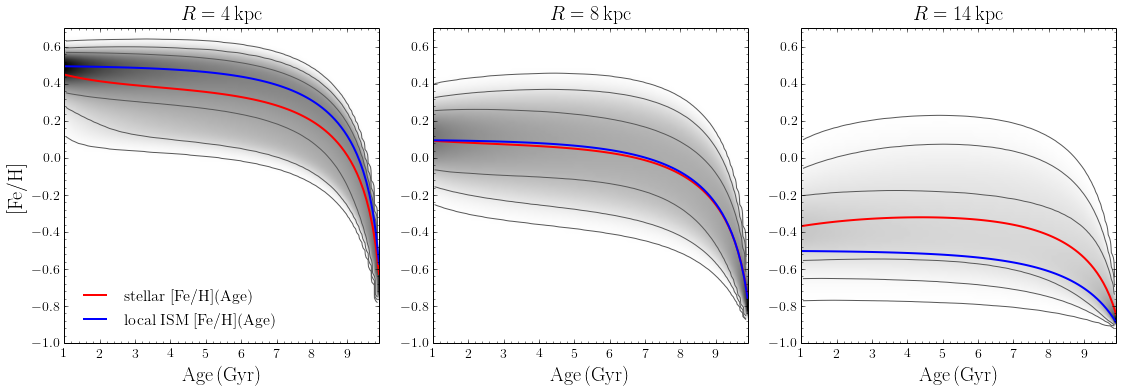}
\caption{The age--metallicity relation at different radii for our
  simple model. The blue line shows the metallicity of the ISM
  as a function of age at different radii, while the solid black lines
  denote 1, 2, and 3$\sigma$ of the distribution of stars in the
  age-metallicity plane respectively; the solid red line gives the
  mean metallicity as a function of age. The solar neighborhood and
  the outer disk both display a wide range of metallicities for all
  but the oldest stars.
\vspace{0.35cm}}
\label{agemetchurn}
\end{figure*}

In order to determine the MDF due to churning, we must also
integrate over the age distribution at a given radius and therefore
need (a) the metallicity as a function of age at a given initial
radius and (b) the age distribution at every $R$. For the former, we
assume that metallicity increases logarithmically with age as a
function of radius, starting at [Fe/H]$=-0.9$ and up to the peak of
the skew-normal intrinsic MDF plus its dispersion; this relation is shown for a
few radii in Figure \ref{agemetchurn}. We approximate
the final age distribution at each radius as the initial age
distribution, which is approximately the case in a more detailed
calculation that takes the effects of churning into account. The
initial age distribution $p(\tau|R)$ at each radius is simply a
consequence of the assumed initial MDF and initial age--metallicity
relation.

\begin{figure}[t!]
\centering
\includegraphics[width=3.0in]{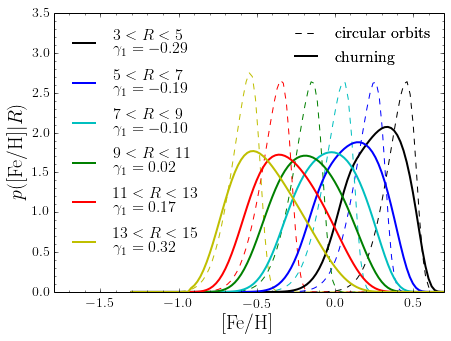}
\caption{The MDF as a function of $R$ with the inclusion of blurring
  and churning. The dashed lines show the initial MDF and the solid
  lines display the MDF including the effects of churning and
  blurring. The redistribution of guiding radii due to churning is
  able to significantly change the skewness compared to the initial MDFs
  and can explain the changes in skewness and its sign observed in the
  APOGEE sample.}
\label{churning}
\end{figure}

We obtain the churning MDF as
\begin{equation}\nonumber
  p([\mathrm{Fe/H}]|R) = \int \mathrm{d} R_g p_c(R_g|R,\tau)\,p(\tau|R)/\left|\mathrm{d}[\mathrm{Fe/H}]/\mathrm{d}\tau\right|\,,
\end{equation}
where the age $\tau$ is a function of $R_g$ and [Fe/H]. We
evaluate this expression at different radii and obtain MDFs that
include the effects of both blurring and churning, displayed in
Figure \ref{churning}. With the addition of churning, we
are able to reproduce the change in skewness observed in the MDFs
across the plane of the disk and in particular the change in sign
around $R=9$ kpc. Stars need to migrate at least 6 kpc to
the outer disk and at least 3 kpc around the solar neighborhood to produce
the observed change in skewness.

The model used in this section is a very simple model, and the predictions from the model do not match our observations perfectly; more detailed and realistic modeling is required that consistently takes into account the chemo-dynamical evolution of the disk. However, these tests demonstrate that blurring alone, as suggested by \citet{Snaith2014a}, is unable to reproduce our observations, and that the addition of churning to our model yields significantly better agreement with the observed MDF, in particular the change in skewness with radius. Therefore, the APOGEE MDFs indicate that migration is of global importance in the evolution of the disk.

In principle, the gas can undergo radial mixing as well, causing enrichment (or dilution) of the ISM at other locations in the Galaxy through processes such as outflows and Galactic fountains. However, in this case the gas must migrate before it forms stars, so the timescales are much shorter. Additionally, many processes that would cause gas to migrate, such as non-axisymmetric perturbations, will also induce stellar migration and can therefore not be decoupled from the stars. It is likely that a combination of both gas and stellar migration is required to reproduce our observations in chemo-dynamical models for the Milky Way.

Finally, we can also calculate the predicted age--metallicity relation at different locations throughout the disk from this model:
\begin{equation}\nonumber
p(\tau,[\mathrm{Fe/H}]|R) = p(R_g|R,\tau)\,p(\tau|R)/\left|\mathrm{d}[\mathrm{Fe/H}]/\mathrm{d}R_g\right|\,,
\end{equation}
where $R_g$ is the initial guiding radius corresponding to $\tau$ and
[Fe/H]. The age--metallicity relation from the model is shown in
Figure \ref{agemetchurn}. This relation is in qualitative agreement with
observations of the age--metallicity relationship of the solar
neighborhood (e.g.,
\citealt{Nordstrom2004,Haywood2013,Bergemann2014}). In our model, the outer disk has a wide range of metallicities at any age due to radial migration, in agreement with models from \citet{Minchev2014a}.

\subsection{Vertical Structure: Heating vs. Cooling}

Our measurements confirm the well established trend of enhanced
\af in stars at greater distances from the plane, and they
show that this trend with $|z|$ holds over radii $3<R<11$ kpc.
At still larger radii, the fraction of high-\af stars
is small at all values of $|z|$. The fact that stars with thin disk chemistry can be found at large distances about the plane in the outer disk can be explained by viewing the disk as composed of embedded flared mono-age populations, as suggested by \citet{Minchev2015}.

The trend of \af with $|z|$ is particularly striking in the
$3-5$ kpc annulus, where the stars lie along the sequence expected
for a single evolutionary track but the dominant locus of stars shifts
from low-\mh, high-\af at $1<|z|<2$ kpc to
high-\mh, low-\af at $0<|z|<0.5$ kpc.
Along an evolutionary track, both \mh and \af are proxies for age.

Heating of stellar populations by encounters with molecular clouds,
spiral arms, or other perturbations will naturally
increase the fraction of older stars at greater heights
above the plane simply because they have more time to experience heating.
However, even more than previous studies of the solar neighborhood,
the transition that we find is remarkably sharp, with almost
no low-\af stars at $|z|>1$ kpc and a completely dominant
population of low-\af stars in the midplane.
Explaining this sharp dichotomy is a challenge for any model
that relies on continuous heating of an initially cold stellar
population.  One alternative is a discrete heating event
associated with a satellite merger or other dynamical encounter
that occurred before SNIa enrichment produced decreasing \af ratios.
Another alternative is ``upside down'' evolution, in which
the scale height of the star-forming gas layer decreases with time, in combination with flaring of mono-age populations with radius (e.g., \citet{Minchev2015}),
as the level of turbulence associated with vigorous star
formation decreases (e.g., \citealt{Bournaud2009,Krumholz2012,Bird2013}).
% JBird, OTHER REFERENCES TO SUGGEST HERE?
In this scenario, the absence of low-\af stars far from the
plane implies that the timescale for the vertical compression
of the disk must be comparable to the $1-2$ Gyr timescale of SNIa
enrichment.  This timescale appears at least roughly consistent
with the predictions of cosmological simulations
(e.g., Figure~18 of \citealt{Bird2013}).
More generally, the measurements presented here of the
spatial dependence of \af-\mh distributions and MDFs
provide a multitude of stringent tests for models of
the formation and the radial and vertical evolution of the
Milky Way disk.

\section{Conclusions}

The solar neighborhood MDF has proven itself a linchpin of Galactic
astronomy, enabling major advances in our understanding of chemical
evolution (e.g., \citealt{VandenBergh1962}) in conjunction with stellar
dynamics \citep{Schonrich2009a}. Galaxy formation models must ultimately
reproduce the empirical MDF as well \citep{Larson1998}. In this paper, we
report the MDF and alpha-abundance as measured by 69,919 red giant stars observed by APOGEE across much of the disk ($3<R<15$ kpc, $0<|z|<2$ kpc). Our simple dynamical model
reveals the exciting prospect that the detailed shape of the MDF is
likely a function of the dynamical history of the Galaxy. Our conclusions are as follows: 

%We use 69,919 red giant stars from DR12 of SDSS-III/APOGEE to measure metallicity distribution functions and characterize the \af vs. \mh plane across the Milky Way disk. 

\begin{itemize}
  \item The inner and outer disk have very different stellar distributions in the \af vs. \mh plane. The inner disk is well characterized by a single sequence, starting at low metallicities and high-\af, and ending at solar-\af abundances with \mh$\sim+0.5$, while the outer disk lacks high-\af stars and is comprised of primarily solar-\af stars.
  \item The scale-height of the inner Galaxy decreased with time, as the (older) metal-poor high-\af stars have large vertical scale-heights, while (younger) metal-rich solar-\af populations are confined to the midplane.
  %\item The MDF of the inner Galaxy is similar in shape to closed-box chemical evolution models, and does not appear to suffer from the G-dwarf problem
  \item The peak metallicity and skewness of the MDF is a function of location within the Galaxy: close to the plane, the inner disk has a super-solar metallicity peak with a negative skewness, while the outer disk is peaked at sub-solar metallicities and has a positive skewness.
  \item Models of the MDF as a function of $R$ that include blurring are unable to reproduce the observed change in skewness of the MDF with location.
  \item Models with migration included match our observations of the change in skewness; migration is likely to be an important mechanism in the observed structure of the disk.
\end{itemize}

M.R.H. and J.H. acknowledge support from NSF Grant AST-1109718, J.B. acknowledges support from a John N. Bahcall Fellowship and the W.M. Keck Foundation, D.L.N. was supported by a McLaughlin Fellowship at the University of Michigan,  J.C.B. acknowledges the support of the Vanderbilt Office of the Provost through the Vanderbilt Initiative in Data-intensive Astrophysics (VIDA), D.H.W, B.A. and J.A.J. received partial support from NSF AST-1211853,  S.R.M. acknowledges support from NSF Grant AST-1109718, T.C.B. acknowledges partial support for this work from grants PHY 08-22648; Physics Frontier Center/{}Joint Institute for Nuclear Astrophysics (JINA), and PHY 14-30152; Physics Frontier Center/{}JINA Center for the Evolution of the Elements (JINA-CEE), awarded by the US National Science Foundation. D.A.G.H. and O.Z. acknowledge support provided by the Spanish Ministry of Economy and Competitiveness under grant AYA-2011-27754.

Funding for SDSS-III has been provided by the Alfred P. Sloan Foundation, the Participating Institutions, the National Science Foundation, and the U.S. Department of Energy Office of Science. The SDSS-III web site is http://www.sdss3.org/. SDSS-III is managed by the Astrophysical Research Consortium for the Participating Institutions of the SDSS-III Collaboration including the University of Arizona, the Brazilian Participation Group, Brookhaven National Laboratory, Carnegie Mellon University, University of Florida, the French Participation Group, the German Participation Group, Harvard University, the Instituto de Astrofisica de Canarias, the Michigan State/Notre Dame/JINA Participation Group, Johns Hopkins University, Lawrence Berkeley National Laboratory, Max Planck Institute for Astrophysics, Max Planck Institute for Extraterrestrial Physics, New Mexico State University, New York University, Ohio State University, Pennsylvania State University, University of Portsmouth, Princeton University, the Spanish Participation Group, University of Tokyo, University of Utah, Vanderbilt University, University of Virginia, University of Washington, and Yale University. 

\bibliographystyle{apj}
\bibliography{ref}

\appendix

\section{Potential Metallicity Biases}

We consider several potential sources of bias in our observed metallicities and their distributions:
\begin{itemize}
  \item Bias from sample selection: color cut and magnitude limits
  \item How well a selection of giants is able to reproduce the underlying MDF of the sampled population
  \item ASPCAP spectral grid edges: spectral library has no results for T$_{\mathrm{eff}}<3500~$K and $\log{g}<0$
  \item Distance Errors
\end{itemize}

\subsection{Correcting For Biases From Sample and ASPCAP Selection}
%APOGEE focuses on observing giants across the disk of the Milky Way, and giants may not be as representative of the underlying populations as other tracers, such as G-dwarfs. 
We use TRILEGAL simulations \citep{Girardi2005} of stellar counts across the Galaxy to determine the magnitude of any sample selection effects on our results.  These simulations provide an opportunity to test how well our targeted sample is able to trace the underlying population of giants as a function of location within the disk.  Note that giants are not a perfect representation of the underlying populations of stars, having a bias against older (generally more metal-poor) stars, with the relative population sampling being roughly proportional to $\tau^{-0.6}$. However, correcting for this bias is difficult, and depends strongly on the star formation histories at different locations throughout the disk. For this work, we do not correct for the uncertain, non-uniform sampling of giants and therefore our MDFs are slightly biased against older populations.

\begin{figure*}[t!]
\centering
\includegraphics[width=6.2in]{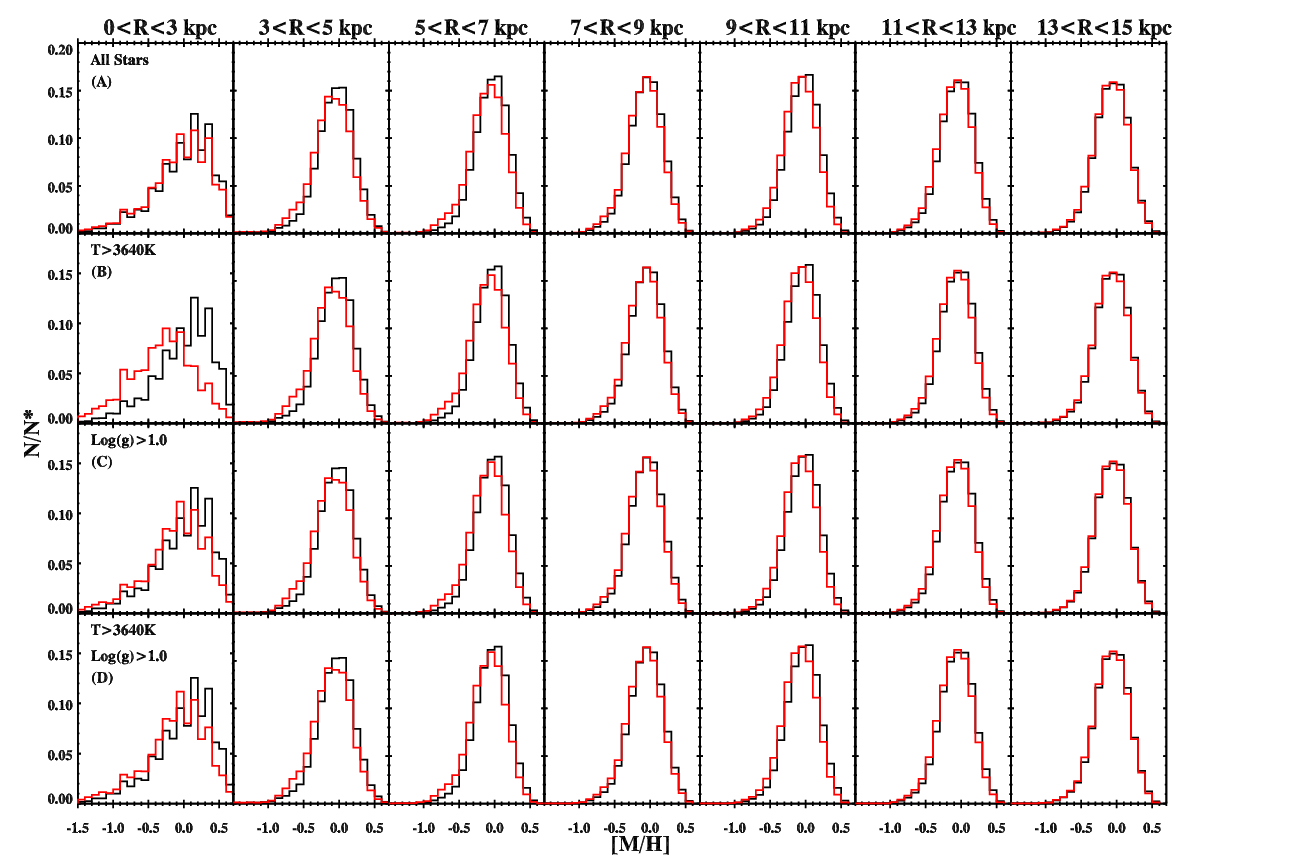}
\caption{The observed mock MDF of giants (red) compared to the intrinsic MDF of giants (black) as a function of $R$. \textbf{(A):} Only targeting criteria (field placement, color cuts, magnitude limits) applied to the input simulations. \textbf{(B):} The observed MDF with the temperature restrictions imposed by the ASPCAP grid edge. \textbf{(C):} The observed MDF with a restricted surface gravity range. \textbf{(D):} The observed MDF with both the temperature and surface gravity restrictions. There is little difference between panels \textbf{(C)} and \textbf{(D)}, the effects of the temperature grid-edge on the observed MDF has largely been mitigated due to the restriction in surface gravity, leading to a significantly less biased MDF.}
\label{trilegalmdf}
\end{figure*}

\begin{figure*}[t!]
\centering
\includegraphics[width=6in]{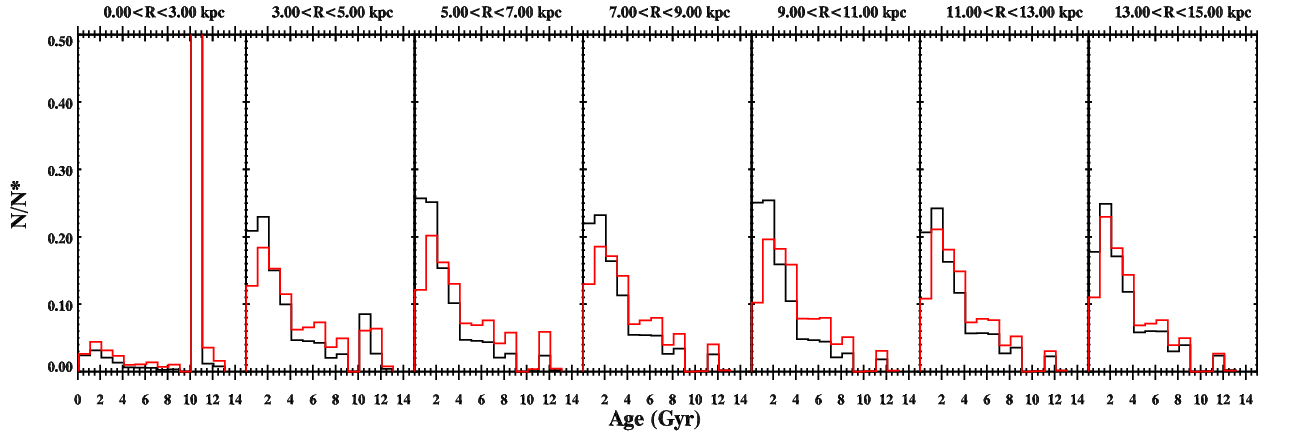}
\caption{The intrinsic age distribution of giants in the TRILEGAL simulation (black) compared to the mock observed age distribution (red). The sample selection criteria does not induce a large bias in the observed sample, although the targeted sample is slightly older in general than the underlying age distribution of giants in the simulation. }
\label{trilegalage}
\end{figure*}

The simulations were run through our targeting software, to obtain an APOGEE-like sample through all of the lines of sight observed in the DR12 dataset. These simulations do not have a radial metallicity gradient, so they are not a perfect representation of our observations, but are still useful for our testing purposes. Because of the lack of a metallicity gradient, the inner disk ($3<R<7$ kpc) is significantly more metal-poor in the simulations than is observed, which means the stars making up the stellar populations in the simulation have higher effective temperatures than would be observed in these locations. Because of this effect, we also include the bulge regions from the simulations ($0<R<3$ kpc) as it is more metal-rich, to aid in our testing of the effects of the ASPCAP temperature grid edge on the observed metallicity distributions. 
The underlying MDF is compared to the observed MDF as a function of radius in Figure \ref{trilegalmdf}. The observed MDF is similar to the underlying MDF based on our sample selection -- we do not have large systematics due to our field selection, magnitude limits, or color cut (panel (A), Figure \ref{trilegalmdf}).  We also compare the ages of the stars we observe to those of the underlying giant populations; the age distribution for our targeted sample of giants is in general older than the underlying sample, but the differences are not dramatic (Figure \ref{trilegalage}). 

However, in the inner Galaxy ($0<R<3$ kpc) there is a significant bias against metal-rich stars due to the temperature edge in the ASPCAP spectral libraries (panel (B), Figure \ref{trilegalmdf}). This result is expected, as the more distant and more metal-rich stars are cooler. The magnitude of this bias on our observed APOGEE sample depends on the number of metal-rich stars present at larger distances, but it is likely to be significant for stars with \mh$>0.1$ based on these simulations. There are several thousand stars in the DR12 catalog that are at the temperature grid edge; most are in fields towards the inner disk and Galactic center, and are likely the most metal-rich stars in the sample.  It is therefore likely that there is a significant bias against metal-rich stars for these fields.

To mitigate this bias, we apply a cut in surface gravity on our sample. Marching down the giant branch to higher surface gravities lowers our sample size, especially at larger distances, but reduces the impact of the temperature grid edge on our sample (demonstrated in Figure \ref{survey}). The surface gravity cut removes the metal-poor, luminous giants that are warmer than the grid edge than the similarly lower surface gravity metal-rich stars, which are removed from the sample due to the temperature restrictions. A cut of $\log{g}>1.0$ was most effective at mitigating the bias due to the temperature grid edge (panels (C) and (D), Figure \ref{trilegalmdf}). There is still a slight bias against metal-rich stars even with this surface gravity cut, but it is significantly less than if no restriction on surface gravity was imposed. It is possible that the observed MDFs for the very inner disk ($3<R<5$ kpc) are slightly more metal-poor than in reality, but the effect is likely not large; the difference in mean metallicity of the MDF from the input simulations and our observed simulations is $\sim0.1$ dex after the surface gravity restriction, compared to $>0.3$ dex due to the effects of the temperature cutoff.

\subsection{Distance Errors}

\begin{figure*}[t!]
\centering
\includegraphics[width=6in]{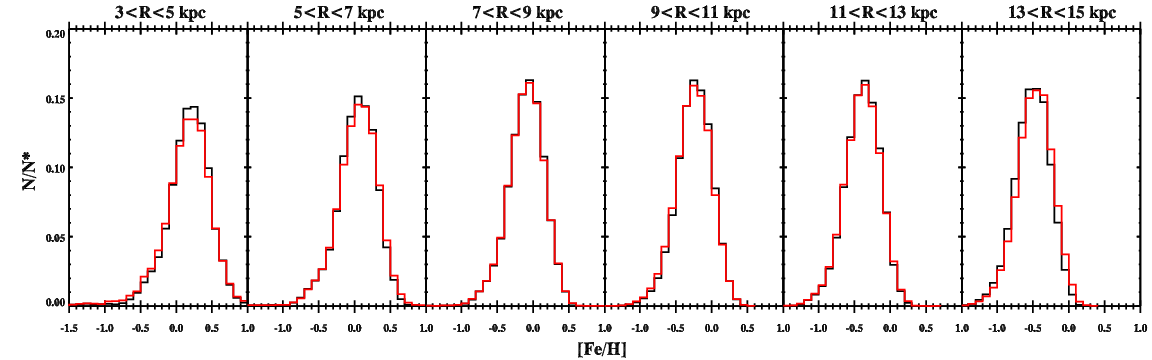}
\caption{The mock targeted MDF (black) versus the mock observed MDF after applying distance errors to the TRILEGAL simulation (red). The difference between the two is negligible except for the very outer disk ($R>13$ kpc).}
\label{distbias}
\end{figure*}

Distances to individual stars are accurate to about 15-20\%. With nearly 70,000 stars in the sample, there are non-negligible amounts of $2$-$3\sigma$ errors in the distances, which could potentially influence the observed metallicity distributions. In particular, the largest distance bins (inner and outer disk) will suffer more from errors in distance calculations, as 20\% errors in distance at these locations is roughly the radial extent of the bin. We use TRILEGAL simulations to test the effect of distance errors on our sample.

A radial gradient of -0.075 dex kpc$^{-1}$ is added to the default MDF of the simulation, such that the inner disk is more metal-rich than the outer disk. Distance errors are modeled using a normal distribution with a dispersion of 20\% and mean of 0. The MDF of the targeted simulations (the ``true'' mock observed MDF) is compared to the  mock MDF with distance errors included in Figure \ref{distbias}. The errors in distance have a negligible effect on the MDF, except for potentially in the very outer disk ($R>13$ kpc). This area of the Galaxy has the fewest stars per $R, |z|$ bins as well, so any scatter from other regions of the disk will have a larger impact on the observed MDF. The change in the MDF due to distance uncertainties even at these larger radii is still small, however, less than $\sim0.1$ dex, so the overall effect on our results as a whole due to distance errors is negligible.

\section{Systematics in \af with Temperature}
There are temperature trends in our \af abundance measurements. Calibrations performed by \citet{Holtzman2015} remove most of these trends, but they lack calibrators for cool stars (T$_{\textrm{eff}}<4000$ K) in most clusters. For stars cooler than this value, there are still temperature trends in \af. The trends in \af with temperature likely reflect that different elements contribute to the overall \af determination at different temperatures. 

\begin{figure*}[t!]
\centering
\includegraphics[width=6.2in]{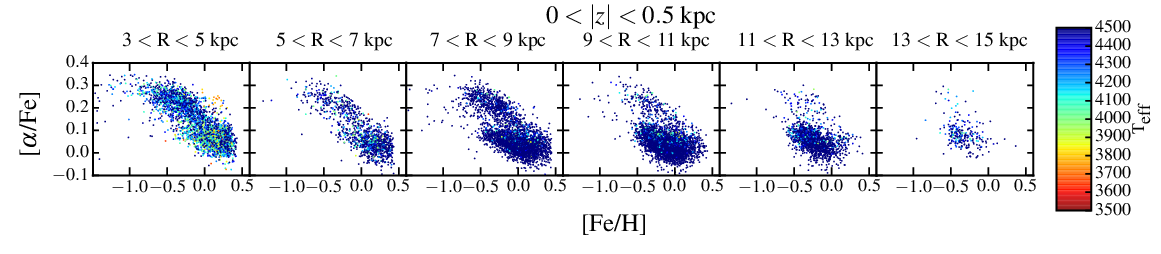}
\caption{The distribution of stars in the \af vs. \mh plane color coded by T$_{\textrm{eff}}$ for stars with $|z|<0.5$ kpc. There is evidence for temperature trends in the \af abundance in the inner Galaxy ($3<R<5$ kpc). This figure is down-sampled and shows only 15\% of the datapoints for $R>5$ kpc.}
\label{alphat}
\end{figure*}

The surface gravity restrictions employed to mitigate the metallicity bias due to the temperature cutoff in the spectral libraries removes most stars that are affected by errors in the \af determination, but some issues remain (Figure \ref{alphat}). The warmer stars have higher \af abundances, while the cooler stars have slightly lower \af abundances. The overall spread due to temperature is small, $\sim0.1$ dex, but is noticeable. It is also possible that the bowl shape of the low-\af sequence observed in the solar neighborhood is due to similar temperature effects: the cooler stars again have lower-\af abundances, blurring the overall shape of the low-\af sequence.

\begin{figure*}[t!]
\centering
\includegraphics[width=6.2in]{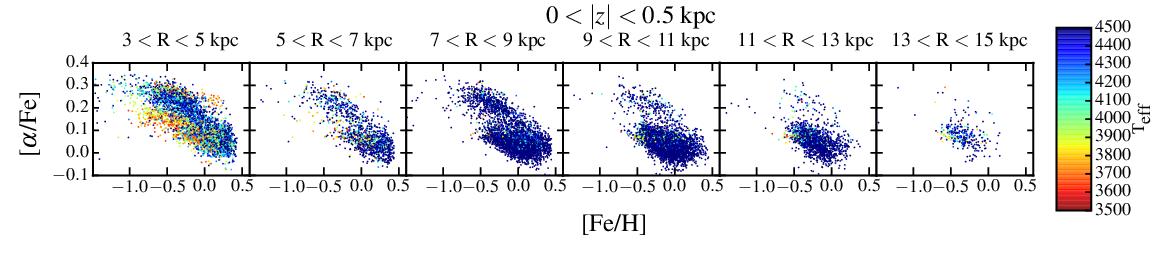}
\caption{Similar to above, but with the surface gravity restrictions removed. The \af abundance trends with temperature are clear with the inclusion of the lower surface gravity stars. Cooler stars have lower \af abundances by $\sim0.1$ dex. }
\label{alphag}
\end{figure*}

\begin{figure*}[t!]
\centering
\includegraphics[width=6.2in]{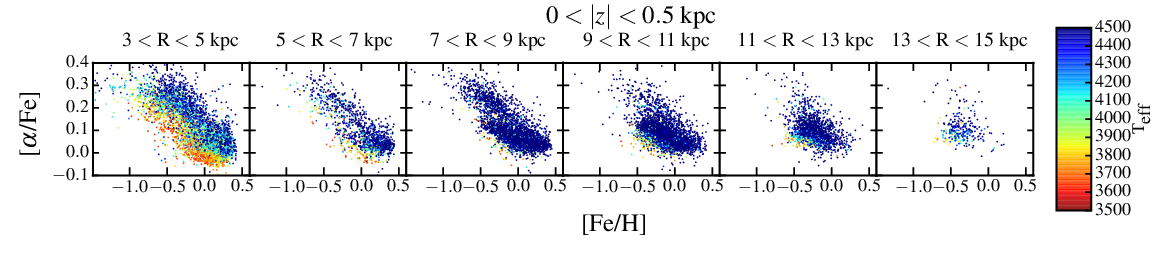}
\caption{Similar to above, but with the \af measured by averaging the abundances for the individual $\alpha$ elements. We find similar trends as in Figure. \ref{alphag} above, with the cooler stars having lower-\af abundances than the warmer stars at the same metallicities.}
\label{manalpha}
\end{figure*}

If the restrictions in surface gravity are not enforced, the appearance of the \af vs. \mh plane is different in the inner Galaxy (Figure \ref{alphag}), and the temperature effects on \af abundance become more apparent. These stars were removed due to the restriction in surface gravity, but are also the coolest stars in the sample that have the most dramatic errors in the \af determination. The fact that there is not a higher surface gravity component to this population is a clue that these stars may have a poor \af determination. We also measure the [$\alpha$/Fe] abundance by averaging over the abundances of the individual $\alpha$ elements (O, Mg, Si, S, Ca, Figure \ref{manalpha}), and find that these stars show a similar temperature gradient compared to the overall \af measurement, with the cooler stars having lower-\af abundances by $\sim0.1$ dex at the same \mh. It is likely that there are errors in the \af abundances for these cool low-surface gravity stars, that are removed by our corrections for the metallicity bias due to the ASPCAP temperature grid edge. It is possible, however, that these abundances are accurate and that there is a second sequence of stars present in the inner disk, this will affect the interpretation of our results for the inner Galaxy.  The \af abundances for these cooler stars are higher than that of the solar-\af populations observed in the rest of the disk, but slightly lower than the high-\af sequence, and if real are a potentially unique stellar population.

\end{document}